\documentclass[aps,pra,twocolumn,superscriptaddress]{revtex4-1}

\usepackage[english]{babel}
\usepackage{amsmath}
\usepackage{amsthm}
\usepackage{amssymb}
\usepackage{mathrsfs}
\usepackage{booktabs}
\usepackage{color}
\usepackage{hyperref}
\usepackage[ruled,vlined]{algorithm2e}
\usepackage{cases}
\usepackage{dblfloatfix}
\usepackage{multirow}
\hypersetup{
  colorlinks   = true,  %Colours links instead of ugly boxes
  urlcolor     = blue,  %Colour for external hyperlinks
  linkcolor    = blue,  %Colour of internal links
  citecolor    = red    %Colour of citations
}
\usepackage{mathtools}
\usepackage{natbib}
\usepackage{pgfplots}
\usepackage{subfigure}
\usepackage{url}

\newcommand{\calH}{\mathcal{H}}
\newcommand{\calL}{\mathcal{L}}

\newcommand{\calX}{\mathcal{X}}
\newcommand{\calZ}{\mathcal{Z}}

\mathtoolsset{showonlyrefs}

\DeclareMathOperator{\Tr}{Tr}

\theoremstyle{definition}

\theoremstyle{remark}

\usepackage{scalerel,stackengine}
\stackMath
\newcommand\reallywidehat[1]{%
\savestack{\tmpbox}{\stretchto{%
  \scaleto{%
    \scalerel*[\widthof{\ensuremath{#1}}]{\kern-.6pt\bigwedge\kern-.6pt}%
    {\rule[-\textheight/2]{1ex}{\textheight}}%WIDTH-LIMITED BIG WEDGE
  }{\textheight}% 
}{0.5ex}}%
\stackon[1pt]{#1}{\tmpbox}%
}
\usepackage{comment}
\usepackage{caption}
\usepackage{float}

\SetArgSty{textnormal}

\makeatletter

\newenvironment{widetext2}{%
  \par\ignorespaces
  \setbox\widetext@top\vbox{%
   \vskip15\p@
   \hb@xt@\hsize{%
    \leaders\hrule\hfil
    \vrule\@height6\p@
   }%
   \vskip6\p@
  }%
  \setbox\widetext@bot\hb@xt@\hsize{%
    \vrule\@depth6\p@
    \leaders\hrule\hfil
  }%
  \onecolumngrid
  \let\set@footnotewidth\set@footnotewidth@ii
}{%
  \par
  \twocolumngrid\global\@ignoretrue
  \@endpetrue
}%

\makeatother

\begin{document}

\title{Pulse-based Variational Quantum Optimal Control for hybrid quantum computing}

\author{R.J.P.T. \surname{de Keijzer}}

\author{O. \surname{Tse}}
% \affiliation{Eindhoven University of Technology, P.~O.~Box 513, 5600 MB Eindhoven, The Netherlands}

\author{S.J.J.M.F. \surname{Kokkelmans}}
% \affiliation{Eindhoven University of Technology, P.~O.~Box 513, 5600 MB Eindhoven, The Netherlands}

\affiliation{Eindhoven University of Technology, P.~O.~Box 513, 5600 MB Eindhoven, The Netherlands}
\affiliation{Eindhoven Hendrik Casimir Institute, P. O. Box 513, 5600 MB Eindhoven, The Netherlands}
\altaffiliation[Corresponding author: ]{r.j.p.t.d.keijzer@tue.nl }

\date{\today}

\begin{abstract}
    This work studies pulse-based variational quantum algorithms (VQAs), which are designed to determine the ground state of a quantum mechanical system by combining classical and quantum hardware. In contrast to more standard gate-based methods, \textit{pulse-based methods} aim to directly optimize the laser pulses interacting with the qubits, instead of using some parametrized gate-based circuit. Using the mathematical formalism of optimal control, these laser pulses are optimized. This method has been used in quantum computing before to design pulses for quantum gate implementations, but has only recently been proposed for full optimization in VQAs \cite{pulsesandbackagain,qocvqe1}. Pulse-based methods have several advantages over gate-based methods, such as faster state preparation, simpler implementation and more freedom in moving through the state space \cite{qocvqe2}. Based on these ideas, we present the development of a variational quantum algorithm employing \textit{adjoint-based} optimal control techniques. This method can be tailored towards and applied in neutral atom quantum computers. Pulse-based \textit{variational quantum optimal control} is able to approximate molecular ground states of simple molecules beyond chemical accuracy. Furthermore, it is able to compete with, or even outperform, the gate-based variational quantum eigensolver in terms of total number of quantum evaluations. The total evolution time $T$ and the form of the control Hamiltonian $H_c$ are important factors in the convergence behaviour to the ground state energy, both having influence on the quantum speed limit and the controllability of the system.
\end{abstract}

\maketitle

\section{Introduction}
\label{sec:section1}

Quantum computing is presently in the noisy intermediate-scale quantum (NISQ) era \cite{Preskill_2018}, where the available quantum computers cannot outperform their classical counterparts.
Nevertheless, even in the NISQ era, quantum computers can be used for specific, well-designed cases \cite{google}. One such application is in solving optimization tasks. In this field, Variational Quantum Algorithms (VQAs) have emerged as the leading optimization technique to obtain quantum advantage on NISQ devices. These algorithms are hybrid, i.e. the algorithm exploits both the classical and quantum computer. Certain VQAs have shown proof of concept for small dimensional problems with various designs of qubits \cite{Kandala,trapion1,photonics1,majorana1,googlehartree}. The common aim of these algorithms is to determine the lowest eigenvalue of a quantum Hamiltonian $H_{\text{mol}}$, which can be equated to finding the ground state energy of a molecule. One example of a VQA is the variational quantum eigensolver (VQE), which evolves an initial state through a gate-based quantum logic circuit to prepare an approximate ground state $|\psi_f\rangle$. In recent years, VQE has been implemented on
many different qubit systems and has become a highly active
area of research (cf.\ \cite{overview1,overview2} for recent comprehensive overviews).

\medskip

Generally, in VQAs, the final qubit states $|\psi_f\rangle$ are prepared from an initial state $|\psi_0\rangle$ by a quantum circuit of parametrized gates. These gates can either manipulate a single qubit or multiple qubits at once, and particularly on a neutral atom qubit system, are implemented using specific laser pulses. The pulses necessary to implement such a gate can be optimized in terms of fidelity, robustness, etc. using the mathematical formalism of optimal control \cite{propson2021robust, Koch, optimizinggates}. This theory finds its use in quantum algorithms such as GRAPE \cite{GRAPE} and CRAB \cite{CRAB}. The entire gate sequence can in principle be seen as a discretization of one continuous laser pulse, as thoroughly explored by Magann et al. \cite{pulsesandbackagain}. From this stems our idea of Variational Quantum Optimal Control (VQOC) as an alternative to VQE. Instead of optimizing a sequence of gate parameters, a laser pulse is optimized to construct a state $|\psi_f\rangle$ close to the ground state of a Hamiltonian $H_{\text{mol}}.$ This idea has been principally investigated by Meitei et al.\ \cite{qocvqe1} and Choquette et al.\ \cite{qocvqe2}. The aim of our work is to explore this option more deeply and systematically by developing a theory for VQOC, and to compare our results with the more established VQE method. The main advantages of our pulse-based method are that the pulses provide more flexibility in the evolution of the initial state, resulting in a broader exploration of the state space and allowing for the total evolution time $T$ to be minimized. Both of these advantages mitigate unwanted effects of qubit lifetimes and decoherence, and are therefore especially important in the NISQ era. 
We further demonstrate how to realistically implement VQOC on a Rydberg atom-based qubit system. From this analysis, we see that VQOC can approximate ground state energies of small molecules beyond chemical accuracy \cite{chemicalaccuracy} on a Rydberg system. 

\medskip

We note that our proposed VQOC algorithm bares similarity to the ctrl-VQE algorithm by Meitei et al.~\cite{qocvqe1}, e.g. pulse-based controls with piecewise constant pulses. However, contrary to ctrl-VQE, the adjoint based approach used in VQOC results in an exact expression of a \textit{continuous time} gradient for the cost functional with respect to the pulses, thus allowing to extend the method beyond square pulses, e.g.\ sum of Gaussians. Furthermore, this approach allows us to prove minimizer existence results, solidifying the well-posedness of the control problem.

\medskip

The layout of this paper is as follows. Sec.~\ref{sec:section2} describes VQAs in the NISQ era. Sec.~\ref{sec:section3} explains the standard VQE method and its shortcomings. In Sec.~\ref{sec:section4} our novel VQOC method is rigorously introduced. In Sec.~\ref{sec:section5} the computational cost on the quantum processing unit of both algorithms is described, as well as our methods for comparing the algorithms. Finally, Sec.~\ref{sec:section6} shows initial results of VQOC, and compares the performance of VQOC to that of VQE.

\section{VQAs in the NISQ era}
\label{sec:section2}
\begin{figure}[H]
    \centering
    \includegraphics[scale=0.33]{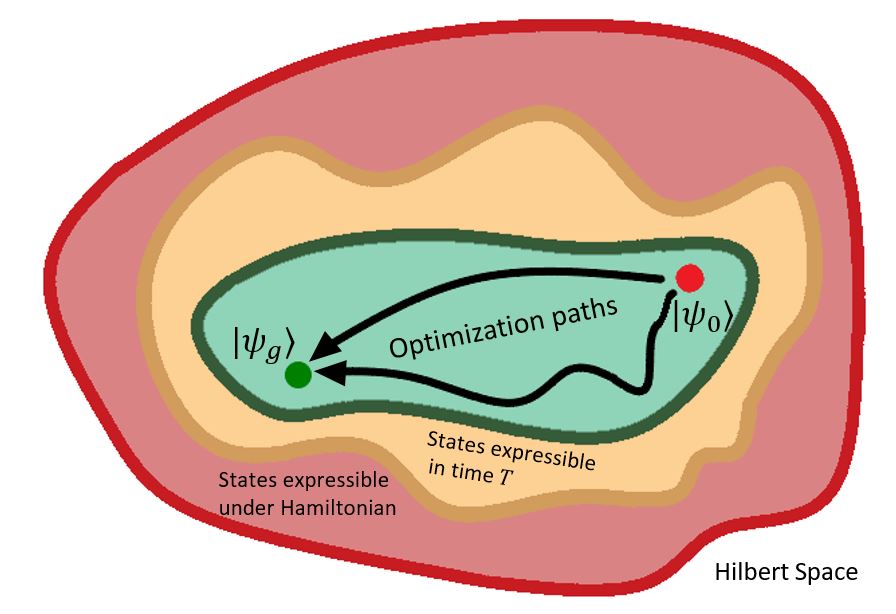}
    \caption{The Hilbert space of qubit states (red) with subsets of the states expressible under the evolution Hamiltonian $H(t)$ (yellow) and the states expressible in total evolution time $T$ (green).}
    \label{fig:hilbertspace}
\end{figure}

The aim of a variational quantum algorithm is to evolve an initial qubit state $|\psi_0\rangle\in \calH^m :=\text{span}(\{|0\rangle,|1\rangle\})^{\otimes m} \simeq \mathbb{C}^{2^m}$ in a given total evolution time $T$, to the
ground state $|\psi_g\rangle$ of a problem Hamiltonian $H_{\text{mol}}\in \calL(\calH^m)$. Here, $\calH^m$ is the Hilbert space of $m$-qubit states, and $\calL(\calH^m)$ is the Hilbert space of operators on $m$-qubits. The Frobenius inner product on $\calL(\calH^m)$ is given by
\begin{equation*}
\langle A, B\rangle_F=\Tr[A^\dagger B].
\end{equation*}
The state evolution, as with every quantum system, is mediated through the Schr\"{o}dinger equation, given in propagator formalism as
\begin{equation}\label{eq:schroedinger}
    t\in(0,T):\quad    i\partial_tU(t)=H(t)U(t);\quad U(0)=I,
\end{equation}
where $U(t)\in\calL(\calH^m)$ is the unitary propagator which describes the evolution of the qubit state as $|\psi(t)\rangle=U(t)|\psi_0\rangle$ \footnote{This work considers  the Hartree unit system in which $\hbar=1$}. The goal of a VQA is thus to manipulate $H(t)$ in a way that makes $U(T)$ map the initial state $|\psi_0\rangle$ to the ground state $|\psi_g\rangle$. To do so, the Hamiltonian $H(t)$ is parametrized, and the variational principle \cite{griffiths} is employed to approximate the ground state. Two important factors in this process are the possible parametrization of the Hamiltonian $H(t)$ and the total evolution time $T$. Figure~\ref{fig:hilbertspace} illustrates the Hilbert space of qubit states. 

\medskip

The possible Hamiltonians $H(t)$ determine a subset of the Hilbert space that is reachable, e.g.\ if the initial state is unentangled and there are no interaction terms in $H(t)$, then all entangled states are unreachable. An even smaller subset is the set of states reachable in a finite time $T$ (given a bound on the infinity norm of $H(t)$). The concept of a minimal time necessary to reach a state $|\psi_f\rangle$ from an initial $|\psi_0\rangle$ is called the \emph{quantum speed limit} (QSL) \cite{qsl1,qsl2,qsl3}. 

Several factors are important in creating a NISQ-friendly VQA: (A) the total evolution time $T$ should be as small as possible to suppress decoherence effects, (B) a large portion of Hilbert space should be expressible, so the algorithm can be used to find the ground states of many $H_{\text{mol}}$, and (C) the evolution should be straightforward to implement and control on the quantum computing system. We will argue in Secs.~\ref{sec:section3} and \ref{sec:section4} that our VQOC method adheres to these criteria better than VQE does.

\section{VQE}
\label{sec:section3}
The variational quantum eigensolver, first proposed in a paper by Peruzzo and McClean in 2014 \cite{firstmention}, is the most widely implemented VQA to date. Using a parametrized quantum logic circuit, a sequence of quantum gates, it has proven able to closely approximate the ground state of several small molecules \cite{overview1,photonics1,trapion1}. See also cf.~\cite{fedorov2021vqe} for a recent overview of VQE.

\medskip

\begin{figure}[H]
    \centering
    \includegraphics[scale=0.95]{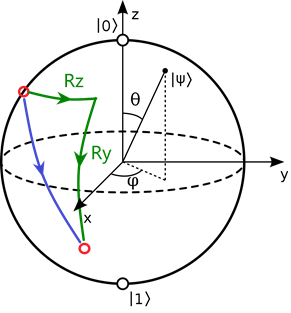}
    \caption{Illustration of NISQ-unfriendliness in VQE. The northern hemisphere state (red dot) needs to be mapped to the southern hemisphere state (red dot). By taking a $R_Z$ gate followed by a $R_Y$ gate, the green path is taken. In terms of technical control, the blue path could have been followed, resulting in faster evolution.}
    \label{fig:rotationbloch}
\end{figure}

In the VQE algorithm, the evolution of the initial qubit state is done via qubit gates. These gates can be described as matrix elements $\calL(\calH^n)$, where $1\leq n\leq m$ is the number of qubits the gate acts on. Examples of these gates are $R_X(\theta_x)$ and $R_Z(\theta_z)$ gates, which respectively rotate a qubit around the $x$-axis by an angle $\theta_x$ and $z$-axis by an angle $\theta_z$ on the Bloch sphere (see Fig.~\ref{fig:rotationbloch}). Other examples include entangling gates such as the canonical CNOT gate and the SWAP gate. A sequence of parametrized and unparametrized gates together with a set of parameters $\vec{\theta}$ determines the evolution.

\medskip

One commonly used template for such a gate sequence alternates between blocks of parametrized single qubit gates $U_{i,j}$, with $i$ indicating the qubit and $j$ the block, and an unparametrized entangling gate $U_{ent}$ (which itself could consist of multiple small gates, such as CNOT gates). The final propagator will take the form
\begin{equation}
\label{eq:equationstatepreperation}
\begin{aligned}
    U(\vec{\theta})=&\left[U_{ent}\prod^m_{q=1} U_{q,d}(\vec{\theta}) \right]\cdots\left[U_{ent}\prod^m_{q=1} U_{q,1}(\vec{\theta}) \right],
\end{aligned}
\end{equation}
where the depth $d$ of a state preparation is defined as the number of blocks in the gate sequence. This method is called the hardware-efficient ansatz \cite{ansatzes} and is especially relevant in NISQ machines where single qubit gates $U_{q,d}$ can be implemented with high fidelity, and the passive interaction between the qubits can be used to create entanglement gates $U_{ent}$ \cite{Preskill_2018}. If such an interaction is not present and the qubit system possesses some form of control in entanglement operations, then there is more freedom in choosing what $U_{ent}$ should look like, and this choice is highly influential on the performance of the algorithm \cite{entanglinggateinfluence}. 

\medskip

After state preparation, the energy of the prepared state on $H_{\text{mol}}$ is determined, which can be done efficiently on a quantum computer for Hamiltonians with sparse decompositions \cite{paulidecomp}. Based on these measurement outcomes, the set of parameters is updated in order to create a state with lower energy. In VQE, stochastic approaches are often implemented to update the parameters. One commonly used example is the simultaneous perturbation stochastic approximation (SPSA) method \cite{SPSA,spsacoefficients,Kandala}. This method samples two points in the parameter space close to $\vec{\theta}$ and measures their energies. Based on these results, a step is taken against the direction of an estimated gradient of the energy w.r.t.\ the parameters. This work will employ the SPSA method for VQE as, similar to our own VQOC method, it is a first order method. The total number of quantum evaluations (QE) per parameter update is thus given by
\begin{equation}
\label{eq:costVQE}
\#\text{QE}=2\cdot\#\text{shots},
\end{equation}
where $\#\text{shots}=O(\epsilon^{-2})$ is the number of times a measurement is to be repeated in order to get an $O(\epsilon)$ error in the expectation values. The product of the number of iterations and the number of quantum evaluations per iteration is thus the total number of times a quantum circuit is run on the system. In the NISQ era, where quantum resources are limited, it is important to keep this number as small as possible while desiring a given accuracy. For SPSA-based VQE, two expectation values are necessary per parameter update, hence the factor 2 (see \cite{Crawford2021efficientquantum} for more detail). In this work, expectation values are calculated as products between matrices and vectors instead of estimated via simulated measurements.

\medskip

When looking at the NISQ-friendly criteria of Sec.~\ref{sec:section2} we see that for VQE, the following issues exist: 
\begin{enumerate}
    \item[(A)] The total evolution time $T$ is fully determined by the gate sequence and can not be decreased unless the gate sequence is altered (see Fig.~\ref{fig:rotationbloch}).
    \item[(B)] Even with enough control available, the gate sequence might inhibit the space of reachable states. For example, if the qubits can be fully rotated on the Bloch sphere by the quantum computing system; if only Z-gates were included in the gate sequence, a large part of the Hilbert space becomes inaccessible.
    \item[(C)] The necessary control to create certain gates might be demanding. For instance, in a Rydberg system, complex and time-consuming laser pulses might be necessary to create specific single qubit or entanglement gates. Implementation is, therefore, not straightforward.
\end{enumerate}
Based on these shortcomings in VQE, we believe that VQOC can serve as a more NISQ-friendly VQA.

\begin{figure}[H]
    \centering
    \includegraphics[scale=0.43]{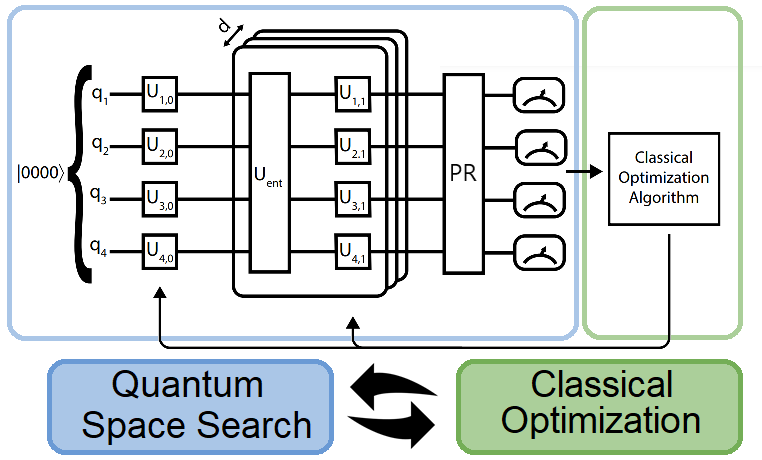}
    \caption{Schematic diagram of VQE illustrating the interplay between quantum space search and classical optimization. An initial state evolves through a hardware-efficient ansatz gate sequence of parametrized gates $U_{i,j}$ and entanglement gates $U_{ent}$. Afterwards, the state is post-rotated (PR) \cite{bravyi1} and measured. A classical optimization algorithm then updates the parameters of the $U_{i,j}$ gates.}
    \label{fig:vqediagram}
\end{figure}

\newpage
\section{Optimal control}
\label{sec:section4}
In this section, the derivation of the pulse-based variational quantum optimal control algorithm (VQOC) as a VQA is given. The motivation for a pulse-based method stems from the idea that every gate is implemented as a laser pulse. Therefore, the entire gate sequence can be seen as a discretization of one continuous laser pulse, as principally explored by Magann et al.\ \cite{pulsesandbackagain}. On the other hand, the origin of gate-based quantum logic circuits stems from an analogy to digital computing. In principle, however, there is nothing digital about the outset of VQAs. Therefore, there is also no a priori reason as to why gate-based algorithms would be preferable.

\medskip
Quantum optimal control theory has been connected to variational state preparation in several works \cite{QSLpulse,qocvqe1,pulsesandbackagain} over the last few years. Recent work has shown that pulse-based methods can indeed prepare desired states in the minimum evolution time set by the quantum speed limit \cite{QSLpulse}. Furthermore, constructing specific control pulses for gate designs focusing on high-fidelity \cite{gatepulse1,gatepulse2,gatepulse3} and robust \cite{gaterobust1, gaterobust2} gates have been a lively topic of research.

\medskip

Our VQOC algorithm can be used to optimize various types of control functions in the qubit system. In particular, this algorithm can optimize the controllable laser pulses that interact with individual qubits. Because of this, we refer to the control functions as pulses. The novelty in our approach lies in the use of an adjoint equation in computing the gradient of the energy with respect to the pulses, which provides a descent direction in the parameter space. A similar approach has been taken by Zhu et al. \cite{rabitzelectric} to specifically optimize electric fields for atom state preparation. Furthermore, we propose a way to compare the results of pulse-based algorithms to gate-based algorithms in terms of energy convergence and time spent calculating on the quantum system, as described in Sec.~\ref{sec:comparing}.

\medskip

To find the ground state energy of a quantum Hamiltonian $H_{\text{mol}}$ with the optimal control formalism, we will need to define the set of {\em admissible pulses}. Let $\calZ:=L^2((0,T);\mathbb{C}^L)$, $L\in\mathbb{N}$, be the space of squared-integrable $L$-dimensional complex-valued functions. For a fixed constant $z_{\max}>0$, we set
\begin{equation*}
\calZ_{ad}:=\bigg\{ z\in \calZ\;\Big|\;\sup_{t\in[0,T]}|z_l(t)|\le z_{\max}\;\; l=1,\ldots,L\bigg\}.
\label{eq:infinitynorm}
\end{equation*}\\
The space of admissible pulses $\calZ_{ad}$ ensures that the control pulses remain physically realizable. The inner product in $\calZ$ is given by
\begin{equation*}
    \langle \tilde{z},z\rangle_\calZ:=\sum_{l=1}^L\int_{0}^T \text{Re}\left[\tilde{z}_l(t)\overline{z_l(t)}\right]\,dt,
\end{equation*}\\
with $\overline{z}$ being the complex conjugate of $z$.

\medskip

We then consider a parametrization of the Hamiltonian $H(t)$ in Eq.~\eqref{eq:schroedinger} of the form
\begin{align}\label{eq:hamiltoniaoc}
    H[z(t)] = H_d + H_c[z(t)],\qquad z\in\calZ_{ad},
\end{align}
where $H_d$ is a drift Hamiltonian representing the passive evolution of the system (see Eq.~\eqref{eq:rydbergvdwinteraction} below), and 
\begin{equation}
\label{eq:controlhamiltonian4}
   H_c[z] := \sum_{l=1}^L \left[ Q_lz_l+Q_l^\dagger\overline{z_l}\right]
\end{equation}
is the control Hamiltonian with control operators $Q_l$, which are not necessarily Hermitian. Examples of what these control operators could represent are discussed in Sec.~\ref{sec:section5}. In the Supplementary Material, we show that the Schr\"odinger equation of Eq.~\eqref{eq:schroedinger} has well-defined solutions for all admissible pulses.

\medskip

The goal of our optimal control problem setup is then to minimize a given real-valued {\em cost functional} $J=J(U,z)$, where $U=U(t)$ are time-dependent unitary propagators satisfying the Schr\"odinger equation \eqref{eq:schroedinger} with prescribed pulses $z\in\calZ_{ad}$. For an extensive description of our optimal control theory, we refer to the Supplementary Material. For the ground state energy problem, we consider the cost functional $J(U,z)=J_1(U) + J_2(z)$, with $J_1$ and $J_2$ given by
\begin{align*}
    J_1(U) &:= \langle\psi(T)|H_{\text{mol}}|\psi(T)\rangle,\quad |\psi(t)\rangle = U(t)|\psi_0\rangle;\\
    J_2(z) &:= \frac{\lambda}{2}\langle z,z\rangle_{\calZ},\quad \lambda>0,
\end{align*} where $|\psi_0\rangle$ is a fixed initial qubit state. The functional $J_1$, therefore, describes the energy of the state generated by the unitary $U$ at time $t=T$, while the $J_2$ term punishes for high amplitude pulses.

\medskip

Altogether, we have a well-defined ground state energy minimization problem, which is shown to admit a minimizer $z_*\in\calZ_{ad}$ in the Supplementary Material. In practice, the parameter $\lambda>0$ in $J_2$ is chosen to be large, in which case the constraint on the pulses does not take effect. In this case, the optimality conditions for the minimization problem read as follows (the derivations are provided in the Supplementary Material):

\begin{widetext}
\begin{equation}\label{eq:U00_vs_D}
\begin{aligned}
    \text{(state equation)}:\qquad &i\partial_t U(t)=H[z(t)]U(t), & U(0)&=I,\\
    \text{(adjoint equation)}:\qquad & i\partial_tP(t)=H[z(t)]P(t), & P(T)&=-2iH_{\text{mol}}|\psi(T)\rangle\langle \psi_0|,\\
    \text{(control equation)}:\qquad &\lambda z_l(t)-\Tr\Bigl[Q_l^\dagger\Bigl(P(t)U^\dagger(t) + U(t)P^\dagger(t)\Bigr)\Bigr]
    =0,& l&=1,...,L.
\end{aligned}
\end{equation}
\end{widetext}

\newpage
\medskip
\newpage
It can be verified that the solution $P\in \calX$ to the adjoint equation is given explicitly in terms of $U$ as
\begin{equation}\label{eq:adjointsolution}
   P(t) =-2iU(t)U^\dagger(T) H_{\text{mol}}|\psi(T)\rangle\langle \psi_0|,
\end{equation}
which is fully dependent on $U$ and the initial parameters. This allows for the second term in the control equation to be expressed as
\begin{equation*}
\begin{aligned}
    \eta_l(t)&:= -\Tr\Bigl[Q_l^\dagger\Bigl(P(t)U^\dagger(t) + U(t)P^\dagger(t)\Bigr)\Bigr]\\
    &= \sum_{k=1}^K-2i\, \left\langle \psi(t)\left| \Big[V_{k,l}^\dagger,\, \Gamma^\dagger(T,t) H_{\text{mol}} \Gamma(T,t)\Bigr]\right|\psi(t)\right\rangle,
    \label{eq:470}
\end{aligned}
\end{equation*}
where $\Gamma(t,s):=U(t)U^\dagger(s)$ describes the evolution by the pulses from $s$ to $t$, $[A,B]=AB - BA$ is the usual commutator, and $V_{k,l}$ are unitaries decomposing $Q_l$ as
\begin{equation*}
Q_l=\sum_{k=1}^K V_{k,l},\qquad l=1,...,L,
\end{equation*}
where $K\in\mathbb{N}$ is the necessary number of unitaries. For practical purposes, where the control terms work on only 1 (or occasionally 2) qubits $K=O(1)$. These terms can be efficiently determined on a quantum computer by first applying the pulse until time $t$, performing a single gate operation based on the parameter gradient algorithm \cite{schuld} (see Supplementary Material), then applying the rest of the pulse up to $T$ and finally measuring the expectation of $H_{\text{mol}}$. Experimentally, this is feasible as long as the gate operation can be done quickly with respect to the drift of the system (e.g. $\tau_V\gg\tau_g$, see Sec.~\ref{sec:section5}). Moreover, the implementation of the pulses will be exactly similar to the previous state preparation except for the one added gate, thus the experimental architecture remains largely unchanged.

\medskip

In order to optimize for $z\in\calZ_{ad}$, we employ the gradient descent algorithm with learning rate $\alpha_k>0$ chosen by Armijo step rule \cite{armijo}:
\begin{equation*}
\begin{aligned}
    z_l^{(k+1)}(t)&=z_l^{(k)}(t)-\alpha_k\bigl(\lambda z_l^{(k)}(t) + \eta_l^{(k)}(t)\bigr),\\
    z_l^{(0)}&\in[-V,V]\cdot10^{-3},
\end{aligned}
\end{equation*}
where the $z_l^{(0)}$ are initialized as low absolute valued constant functions on the interval $[0,T]$. In order to implement the algorithm, the pulses are discretized as equidistant piecewise constant functions with $N=T/\tau\in\mathbb{N}$ steps. For this discretization, the propagator can be expressed in closed form. Let $t_n:= n\tau$, $n=0,\ldots,N-1$, and set $z_l(t) := z_{l,n}$ for $t\in [t_n,t_{n+1})$, $l=1,\ldots,L$. Then for $t\in[t_n,t_{n+1})$, the propagator $U$ takes the form
\begin{equation*}
\begin{aligned}
    U(t) &= \exp\left(-i(t-t_n)H[z(t_n)]\right)\cdot\\
    &\qquad\cdot \exp\left(-i\tau H[z(t_{n-1})]\right)\cdots \exp\left(-i\tau H[z(t_0)]\right).
\end{aligned}
\end{equation*}
The minimization procedure is then executed by the algorithm below.

\begin{algorithm}[H]
\SetAlgoLined
\SetInd{0.5em}{0.5em}
\SetKwData{Left}{left}\SetKwData{This}{this}\SetKwData{Up}{up}
\SetKwFunction{Union}{Union}\SetKwFunction{FindCompress}{FindCompress}
\SetKwInOut{Input}{input}\SetKwInOut{Output}{output}
\Input{$z^{(0)}\in \mathbb{C}^{L\times N}$, $H_{\text{mol}}$, $|\psi_0\rangle$, \#iterations}
\Output{$z^{(\text{\#iterations})}, E$}
\BlankLine
\For{$k=0$ \KwTo $ \text{\#iterations}-1$}{
$U=S(z^{(k)})$;\\
    \For{$n=1$ \KwTo $N$}{
    $P_n=-2i\Gamma(t_n,t_N) H_{\text{mol}}\Gamma(t_N,t_n)|\psi_0\rangle \langle\psi_0|$;\\
    $\eta_{l,n}^{(k)}= \Tr[Q_l^\dagger\left(P_nU^\dagger(t_n)+U(t_n)P_n^\dagger\right)]$;\\
        \For{$l=1$ \KwTo $L$}{
        $z^{(k+1)}_{l,n}=z_{l,n}^{(k)}-\alpha_k\bigl(\lambda z_{l,n}^{(k)}-\eta_{l,n}^{(k)}\bigr)$;
        }
    }
}
$U_f=S(z^{(\text{\#iterations})})$;\\
$E=\langle \psi_0|U_f^\dagger(T)H_{\text{mol}}U_f(T)|\psi_0\rangle$;\\
\caption{Discrete pulses VQOC algorithm}\label{algo:QOCcontinuous2}
\end{algorithm}

\medskip

The total number of quantum evaluations necessary in one parameter update is given by
\begin{equation}
\label{eq:costQOC}
    \#\text{QE}=2\cdot K\cdot L\cdot N\cdot \#\text{shots}.
\end{equation}

\section{Rydberg Physics}\label{sec:section5}
This section introduces basic Rydberg physics to identify what drift and control Hamiltonians  as in Eq.~\eqref{eq:hamiltoniaoc} can look like on a Rydberg atom quantum computing system. This will also yield candidates for the control operators $Q_l$, as in Eq.~\eqref{eq:controlhamiltonian4}.

\medskip

The following analysis focuses on a ground-Rydberg qubit system \cite{overview1} (or Rydberg-Rydberg in the case of dipole-dipole interactions). Because the qubits never leave the $\calH^m$ manifold, only these states are taken into account (assuming that the system evolution time $T$ is well below the lifetime of the states).
 \medskip

The Rydberg states have a passive `always-on' interaction, which is described using a drift Hamiltonian $H_d$, depending on the choice of qubits scheme \cite{rydbergqubit}, as a Van der Waals interaction \cite{vdwaals} or a dipole interaction \cite{dipole}
\begin{equation}
\begin{aligned}
    H_{d,\text{VdW}}&=\sum_{i=1}^m\sum_{j=1}^m\frac{C_6}{R_{ij}^6}|11\rangle_{ij}\langle 11|_{ij},\\
    H_{d,\text{dip}}&=\sum_{i=1}^m\sum_{j=1}^m\frac{C_3}{R_{ij}^3}\left(|01\rangle_{ij}\langle 10|_{ij}+\text{h.c.}\right),
    \label{eq:rydbergvdwinteraction}
    \end{aligned}
\end{equation}
where $R_{ij}$ is the interatomic distance. In this work, the qubit are arranged in a line with equal nearest neighbour distance, such that $R_{ij}=R|i-j|$,  with $R$ being the nearest neighbour distance.

\medskip

The coupling strength and the interaction strength are respectively characterized by the Rabi frequency $\Omega_R$ and the interaction strength $V=C_i/R^i, i\in\{3,6\}$. This leads to two characteristic timescales $\tau_g=1/\Omega_R$, characterizing single qubit manipulation time, and $\tau_V=1/V$, characterizing entanglement time.

\medskip

To perform single qubit manipulations on qubit $j$, a nearly monochromatic laser interacts with the atom to realize the Hamiltonian \cite{rydbergqubit,rydberg1,rydberg2}
\begin{equation}
\begin{aligned}
    H_{j}^{01}(t)&=\frac{\Omega_{j}(t)}{2} e^{i \varphi_{j}(t)}|0\rangle_j\langle 1|_{j}+\text{ h.c.}\\
    &\qquad\qquad-\Delta_{j}(t)|1\rangle_{j}\langle 1|_j,
\end{aligned}
    \label{eq:qubitlightinteraction}
\end{equation}
where `h.c.'\ refers to the hermitian conjugate of the term that precedes it. Here, $\Omega_j(t)$ [Hz] denotes the coupling strength, $\varphi_j(t)$ the phase of the laser coupled to atom $j$, and $\Delta_j(t)$ = $\omega_j(t)-\omega_0$ [Hz] the detuning of the laser frequency $\omega_j(t)$ from the energy level difference $\omega_0$. In the current state-of-the-art systems, the control over the laser coupling $\Omega_j(t)$ and the detuning $\Delta_j(t)$ is much higher than that of the phase $\varphi_j(t)$ \cite{laserphase}. Therefore, often $\varphi_j(t)=\varphi_j\in[0,2\pi)$ is taken constant. It is however not unimaginable that laser phase control will improve significantly in the coming years.

\medskip

The control Hamiltonian $H_{c}$ can contain several terms depending on which of the mentioned laser parameters can be controlled. In general, the control Hamiltonian will take the form (cf.\ Section~\ref{sec:section4})
\begin{equation}
    H_{c}[z(t)]=\sum_{l=1}^LQ_lz_l(t)+Q_l^\dagger \overline{z_l(t)},
    \label{eq:controlhamiltonian}
\end{equation}
where $z_l(t)\in \mathbb{C}$ [Hz], and $Q_l$ is a $m$-qubit operator. The choice of $z_l(t)$ being a complex number stems from the fact that it represents both the coupling strength and the phase in Eq.~\eqref{eq:qubitlightinteraction}. However, as indicated before, the phase is often not controllable. In that case, $z_l(t)\in\mathbb{R}$ is required. This would result in 
\begin{equation*}
    H_{c}[z]=\sum_{l=1}^L(Q_l+Q_l^\dagger) z_l,
\end{equation*}
where $Q_l+Q_l^\dagger$ is Hermitian. However, choosing $Q_l$ Hermitian from the start gives
\begin{equation*}
    H_{c}[z]=\sum_{l=1}^L Q_l (z_l + \overline{z_l}),
\end{equation*}
where $z_l+\overline{z_l}\in \mathbb{R}$. Thus, if there is no control of the phase of the complex number, the formulation of Eq.~\eqref{eq:controlhamiltonian} remains valid by choosing $Q_l$ Hermitian. 

\medskip

In this part all the imaginable and realistic controls on a Rydberg system are listed, and at every future simulation it is indicated which terms are chosen to be included. For the control Hamiltonian $H_c$
\begin{equation*}
    H_{c}[z]=H_{c}^{\text{coup}}[z^{\text{coup}}]+H_{c}^{\text{det}}[z^{\text{det}}]+H_{c}^{\text{ent}}[z^{\text{ent}}],
\end{equation*}
with $z = (z^{\text{coup}}, z^{\text{det}}, z^{\text{ent}})\in \calZ$, $L=2m + m(m-1)$. Here, the coupling control $H_{c}^{\text{coup}}$ and detuning control $H_{c}^{\text{det}}$ take the form of Eq.~\eqref{eq:controlhamiltonian} with
\[
    Q_l^{\text{coup}}=|0\rangle\langle1|_l\quad\text{and}\quad Q_l^{\text{det}}=|1\rangle\langle1|_l,
\]
respectively. Notice that the coupling Hamiltonian $H_c^{\text{coup}}$, already allows for full control on the Bloch sphere of each individual qubit. This is why this term is also referred to as \textit{rotational control}. The entanglement control for respectively Van der Waals interactions and Dipole-Dipole interaction is given by
\begin{equation}
\begin{aligned}
    H_{c}^{\text{ent}}[z]&=\sum_{l=1}^m  \sum_{k\neq l}^m \frac{1}{R_{lk}^6} \left(z_{lk}+\overline{z_{lk}}\right)|11\rangle_{lk}\langle 11|_{lk},\\
    H_{c}^{\text{ent}}[z]&=\sum_{l=1}^m \sum_{k\neq l}^m \frac{1}{R_{lk}^3} \left(z_{lk}+\overline{z_{lk}}\right)(|01\rangle_{lk}\langle 10|_{lk}+\text{h.c.}).
    \label{eq:entanglementham}
\end{aligned}
\end{equation}

An entanglement control could for instance be realized by tuneable Rydberg dressing \cite{rydberg1}. In many cases, the entanglement controls will be taken equal for all pairwise interactions, such that all $z_{lk}$ are equal for every combination of $l$ and $k$.

\section{Comparing VQOC and VQE}
\label{sec:comparing}
This section outlines our devised method for comparing and contrasting VQOC and VQE. We compare both the expressibility of the algorithms and their performance in variational problems.
\medskip

\subsection*{Equivalent evolution processes}

In order to fairly compare VQOC and VQE, several algorithmic parameters need to be considered to create an \textit{equivalent evolution processes}, meaning that the system evolves under the most similar circumstances for both algorithms. Concretely, this means that the algorithms are processed on the same hardware, with the same Hamiltonian, but with different control of the time-dependent interaction parameters to either do pulse-based or gate-based interactions.

\medskip

First, the drift and control Hamiltonians in VQOC should be linked to the ansatz of VQE. In our simulations, both VQE and VQOC are considered for a gr-qubit \cite{rydbergqubit} Rydberg system, where $|0\rangle$ is a ground state and $|1\rangle$ a Rydberg state. The passive interaction between $|1\rangle$ states is given by either the Van der Waals or dipole interaction Hamiltonian $H_d$, as in Eq.~\eqref{eq:rydbergvdwinteraction} \cite{overview1}. The qubits are considered on a straight line, equidistant from one another. For VQOC only rotational control is considered $(L=m)$. The hardware-efficient ansatz is considered in VQE with alternating layers of single qubit rotations and $m$-qubit entanglement gates $U_{ent}$ as in
Eq.~\eqref{eq:equationstatepreperation}. The assumption $\tau_V\gg \tau_g$ is made so that the hardware-efficient ansatz can physically be realized by taking $U_{ent}=\exp(-iH_d\tau_V)$. Without this assumption, the system would passively evolve considerably while the single qubit manipulations are being executed, thus preventing them to be treated as independent. Each state on the Bloch sphere of a single qubit can be reached with a $ZXZ$-rotation. Such a rotation on a qubit $q$ at a depth $i$ can be written as
\begin{equation}
    U_{q,i}(\vec{\theta})=Z_{\theta_1^{q,i}}X_{\theta_2^{q,i}}Z_{\theta_3^{q,i}},
\end{equation}

\noindent where $\vec{\theta}$ has three elements for every qubit and depth pair. For VQE, these $ZXZ$ rotations are considered on the individual qubits in order to get analogous control to the rotational control for VQOC. We assume a Rabi frequency of $\Omega_R=1$ kHz. This gives a gate time of $\tau_g=1$ ms. Adhering to the condition $\tau_V\gg\tau_g$ can always be realistically done by varying $R$ in $V=C_6/R^6.$ Here we set $V=0.1$ kHz. For the VQE algorithm, $T_{\text{VQE}}=d(\tau_g+\tau_V)$ with $d$ the depth of the hardware-efficient ansatz. To get similar evolutions, we take $T_{\text{VQOC}}=T_{\text{VQE}}$, which completes the construction of the \textit{equivalent evolution processes}, as illustrated in Fig.~\ref{fig:sample}.

\subsection*{Expressibility analysis}

To compare the expressibility of VQE and VQOC, we follow an approach employed by Sim et al. \cite{expressibility}. In their work, the parameters of the evolution process are chosen according to some probability distribution. This results in a distribution over the unitaries. Expressibility of the evolution process is then equated to correspondence with the Haar measure, i.e.\ the uniform measure on unitaries \cite{haarmeasure}. To determine this correspondence, unitaries are drawn from the distributions and applied to an initial state $|\psi_0\rangle$. The fidelity of the prepared state with respect to the initial state $F=|\langle\psi_f|\psi_0\rangle|^2$ is determined, resulting in a distribution over fidelities. The distribution of fidelities of the evolution process is then compared to that of the Haar measure in terms of the Kullback-Leibler (KL) divergence $D_{\mathrm{KL}}$, which is a measure of the similarity between two distributions $P$ and $Q$, given by
\begin{equation}
    D_{\mathrm{KL}}(P \| Q)=\sum_{x \in \mathcal{X}} P(x) \log \left(\frac{P(x)}{Q(x)}\right),
\end{equation}
where $\mathcal{X}$ is a probability space. The distribution of fidelities for Haar random states is given by
\begin{equation}
    P_{N}(F)=(N-1)(1-F)^{N-2},
\end{equation}
where $N$ is the dimensionality of the unitaries \cite{haardistribution}.

We consider a VQE hardware efficient ansatz as described above and the VQOC equivalent evolution process. Initial gate parameters/pulses are then chosen randomly for both VQE and VQOC. In VQE, the rotation angles are chosen uniformly in the interval $[0,2\pi]$. For VQOC the piecewise constant values are chosen uniformly at random between $[-1,1]$ kHz. In the results, it will be shown that even for a low choice of $N$, thus limiting the degrees of freedom in VQOC, the expressibility, in terms of correspondence to the Haar measure, is higher in VQOC than for the equivalent VQE circuit.

\subsection*{Variational problems}

\medskip
Next, the algorithms are quantitively compared for their performance in solving variational problems. A good quantity used to compare two quantum optimization algorithms is the total number of circuits run on the quantum computer, during the algorithm. This holds especially true in the NISQ era, where the availability of quantum computing power is limited. Based on Eqs.~\eqref{eq:costVQE} and \eqref{eq:costQOC}, we find the ratio of number of quantum evaluations for both methods per iteration to be 
\begin{equation}
    \frac{\#\text{QE}_{VQOC}}{\#\text{QE}_{VQE}}=K\cdot L\cdot N.
\end{equation}
Recall that $K$ is the number of decomposition terms in the control factors, $L$ is the number of controls scaling linearly with $m$, and finally, $N=T/\tau$, the number of considered time steps in VQOC. For all simulations in this work, we set $K=2$, $L=m$ for rotational control, and $N=100$. In order to characterize the performance on variational problems, several small molecule ground state energies are approximated using both algorithms for a fixed number of total quantum evaluations.

\medskip

Obviously, many other ansatzes for VQE, as well as discretizations for VQOC, are possible. The hardware-efficient ansatz and piecewise constant discretization are the most common in the literature, and we foresee no reason why other procedures should yield substantially different results. Nonetheless, different implementations and comparisons might yield interesting results in future research.

\begin{widetext2}
    \begin{minipage}[b]{\linewidth}
        \begin{figure}[H]
            \centering
            \includegraphics[scale=0.35]{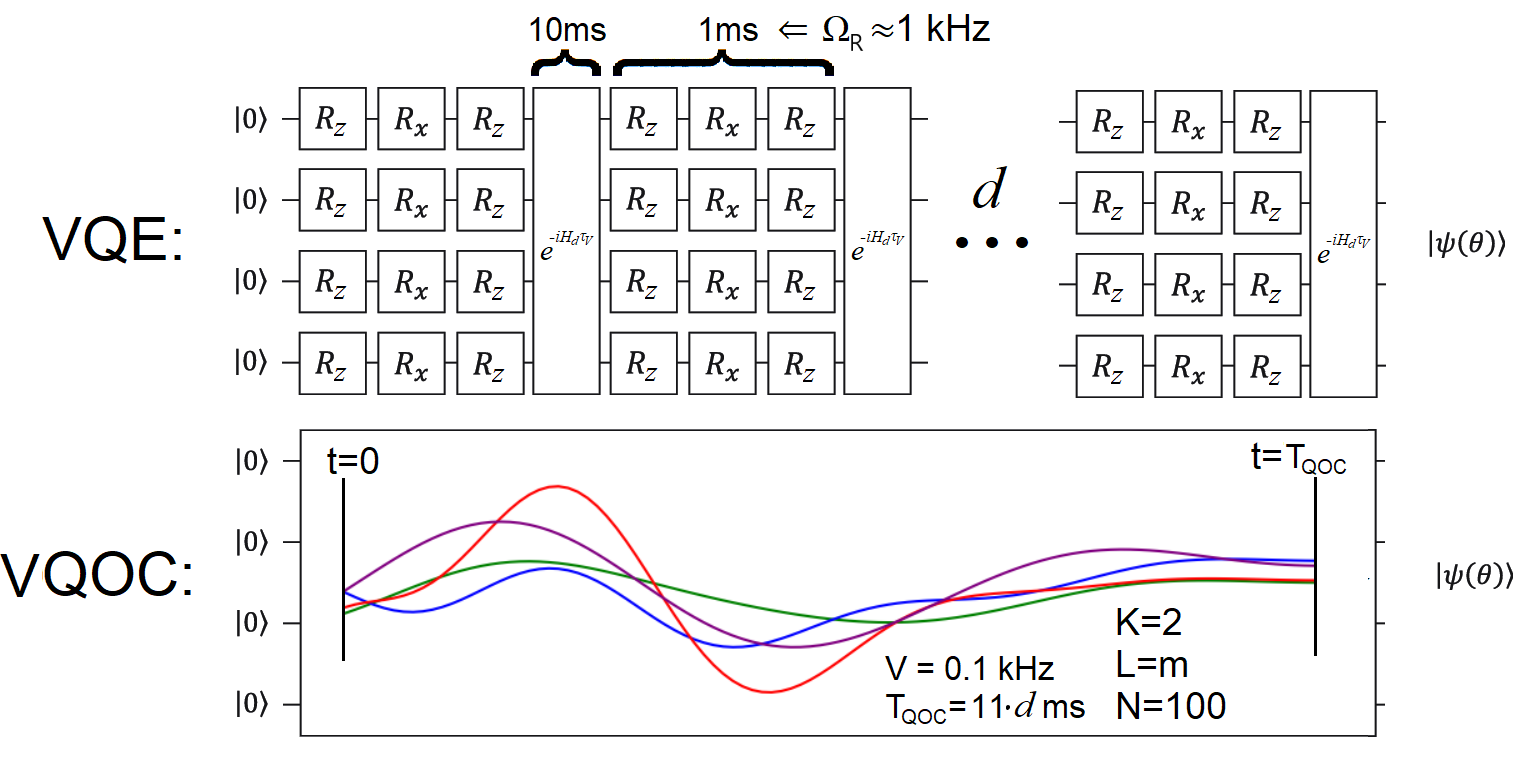}
    \caption{The equivalent evolution processes used to compare VQE and VQOC in the ground-state-energy problem. VQE is conducted on a hardware-efficient ansatz with $d$ layers of alternating $ZXZ$-gates, taking time $\tau_g$, and passive interaction via the drift Hamiltonian, for a time $\tau_V$. VQOC is executed with piecewise constant functions consisting of $N$ steps and $L=m$ rotational controls (giving $K=2$). We assume $\Omega_R=1$ kHz, leading to $\tau_g\approx 1$ ms. Adhering to $\tau_V\gg\tau_g$, we choose $\tau_V=10$ ms which in turn gives $V=0.1$ kHz and $T_{\text{VQE}}=T_{\text{VQOC}}=11d$ ms.}
            \label{fig:sample}
        \end{figure}    
    \end{minipage}
\end{widetext2}

\newpage
\textcolor{white}{.}
\newpage
\textcolor{white}{.}
\newpage

\section{Results}
\label{sec:section6}

\subsection*{Expressibility analysis}
The expressibility analysis, as described in Sec.~\ref{sec:section4}, is performed between VQE and VQOC. Because of finite sampling size, the obtained fidelities are binned together in equally sized bins in the interval $[0,1]$. The KL-divergences are calculated over the obtained histograms.  Fig.~\ref{fig:histograms} depicts the results for $d=2$, $V=0.1$ kHz, $N=4$ and $T_{\text{VQE}}=T_{\text{VQOC}}=22$ ms. Fig.~\hyperlink{fig:histogramsa}{\ref{fig:histograms}a} shows that for $10^4$ samples, the VQOC evolution procedure approximates the Haar measure considerably better than VQE with $d=2$, as it is less concentrated around fidelity $F=0$. Interestingly, this is already significant for the low number of time steps $N=4$. Furthermore, Fig.~\hyperlink{fig:histogramsb}{\ref{fig:histograms}b} shows the Bloch sphere representation of the partially traced first qubit density matrices $\rho_1=\Tr_2[|\psi_f\rangle\langle\psi_f|]$. From these plots, one observes that the density matrices for VQE concentrate more on the poles of the Bloch sphere than for VQOC, indicating less expressibility in terms of correspondence to the Haar measure.  

\begin{minipage}[t]{\linewidth}
\begin{figure}[H]
\begin{centering}
\captionsetup{justification=raggedright}
    \begin{minipage}[t]{.95\textwidth}
         \hypertarget{fig:histogramsa}{}
         \includegraphics[width=\columnwidth]{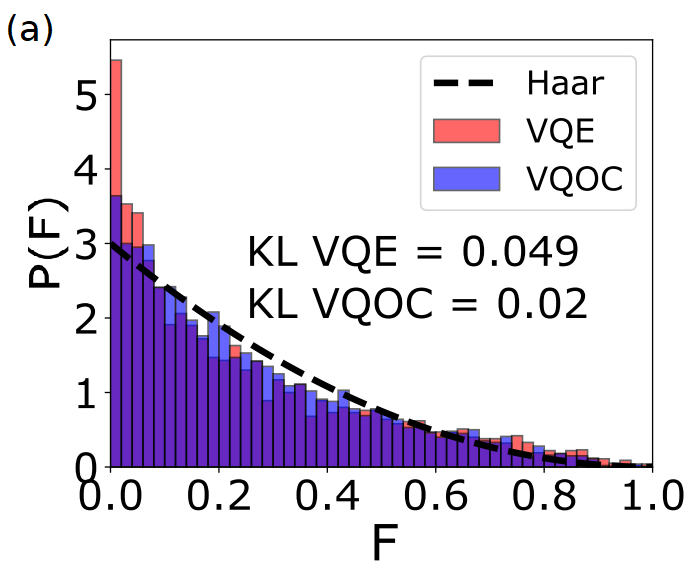}
    \end{minipage}%
    \\
    \begin{minipage}[t]{.99\textwidth}
         \hypertarget{fig:histogramsb}{}
         \includegraphics[width=\columnwidth]{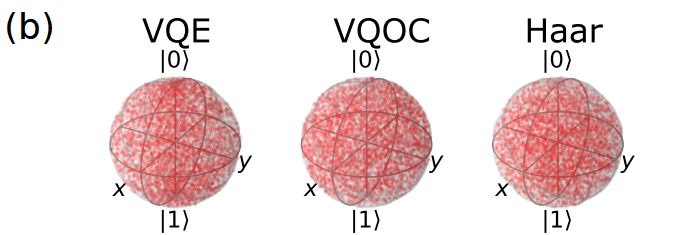}
\end{minipage}%
\caption{Expressibility analysis of VQOC and VQE. VQE with $d=2$, $T_{\text{VQE}}=22$ ms, VQOC with $T_{\text{VQOC}}=22$ ms and $N=4$. (a) Histograms of fidelities between $10^4$ initial and output states, related to the analytic probability distribution of the Haar measure. (b) Representations on the Bloch sphere of the first qubit density matrices $\rho_1=\Tr_2[|\psi_f\rangle\langle\psi_f|]$.}
    \label{fig:histograms}
    \end{centering}
\end{figure}
\end{minipage}

\subsection*{Electronic structure problems}
This work considers ground state problems for H$_2$, LiH and H$_4$ molecules. The H$_2$ and LiH molecules are aligned on the $x$-axis with varying interatomic distance. For H$_2$ the $1s$ orbitals are considered active, higher energy orbitals are ignored \cite{mcquarrie2008quantum}. This results, after spin and electron number reduction \cite{entanglinggateinfluence}, in a 2-qubit problem. For LiH, the $1s$ orbital on the Li atom is fixed as fully occupied. The $2s$ and $2p_x$ orbitals of Li and the $1s$ orbitals of H are considered active. This results in a 4-qubit problem after spin and electron number reduction. For H$_4$, the atoms are aligned on the $x$-axis with all interatomic distances between neighbouring atoms taken equally. Only the 1s orbitals of all 4 H atoms are considered active. This results, after spin and electron number reduction, in a 6-qubit problem \footnote{In this work the molecular Hamiltonians have been determined in the STO-3G basis using quantum computation libraries \textit{OpenFermion} \cite{openfermion} and \textit{Psi4} \cite{psi4}.}.

\medskip

The choice for these molecules as illustrating use cases is based on the fact that all possess weakly and strongly entangled ground states at the interatomic distances considered. For all molecules considered, the exact $H_{\text{mol}}$ ground state energy has been calculated using the FCI method \cite{mcquarrie2008quantum}. In certain simulations, random Hamiltonians are considered. These Hamiltonians are constructed by taking a fixed number of Pauli strings at random and picking weights in the interval $[-1,1]$ uniformly at random. The reason for using these random Hamiltonians is that, by construction, they will possess (with high probability) a highly entangled ground state. This state will also (with high probability) be a linear combination of all basis states, and therefore will have a high QSL w.r.t.\ all basis states.

\medskip

We note that the systems considered in this work are small in terms of the number of qubits. In larger systems, for both gate-based and pulse-based control, a trade-off will have to be made between expressibility and optimisability due to the presence of barren plateaus \cite{rabitztrainability}.

\subsection*{Variational problems}

As a first qualitative illustration of VQOC, the ground states of 2-qubit H$_2$ Hamiltonians (4-qubit LiH Hamiltonians) are calculated for varying interatomic distances, with the vacuum state ($|\psi_0\rangle=|00...0\rangle$) as the initial state. VQOC is run for 100 iterations with rotational control $H_c^{\text{coup}}$, $L=m$, $K=2$, Rydberg interactions as in Eq.~\eqref{eq:rydbergvdwinteraction} with $C_6/R^6=0.1$ kHz, $T=100$ ms and $N=100$ \footnote{In this work all VQOC simulations have been performed using Python Library QuTiP \cite{qutip} while the VQE calculations have been performed using MATLAB R2019b \cite{matlab}.}. The results in Fig.~\ref{fig:H2simulationsQOC} for H$_2$ show that VQOC is able to find energies below the Hartree-Fock energy and reach chemical accuracy \cite{chemicalaccuracy} after as few as 30 iterations. The errors reached are comparable to other (pulse-based) VQAs \cite{qocvqe1,h2results1,h2results2} and well below chemical accuracy. For LiH, all possible basis states ($|00...00\rangle, |00...01\rangle, ... |11...11\rangle$) are taken as initial states, after which the mean and best case energy results are reported, as similarly done in Ref.~\cite{rabitzbestaverage} by Rabitz et al. Fig.~\ref{fig:LIHsimulationsQOC} shows that VQOC is able to get below the Hartree-Fock state energy and reach chemical accuracy \cite{chemicalaccuracy} for LiH after 50 iterations in many occasions.

\medskip

As another qualitative analysis,
Fig.~\ref{fig:LIHsimulationsQOC2} shows final pulses and density matrix evolution for simulations with rotational and entanglement control as in Eq.~\eqref{eq:entanglementham} ($L=m+1, K=2$), Rydberg interactions with $V=0.1$ kHz, $T=100$ ms and $N=100$. It is shown that the density matrices are closely approximated, and the achieved energy errors, which are around $10^{-5}$ Hartree, are well below chemical accuracy. Moreover, the fidelities $F$ of the prepared state with respect to the ground state are close to 1, as desired.

\begin{widetext2}
    \begin{minipage}[b]{\linewidth}
\begin{figure}[H]
\begin{centering}
\captionsetup{justification=raggedright}
    \begin{minipage}[t]{.49\textwidth}
         \includegraphics[width=\columnwidth]{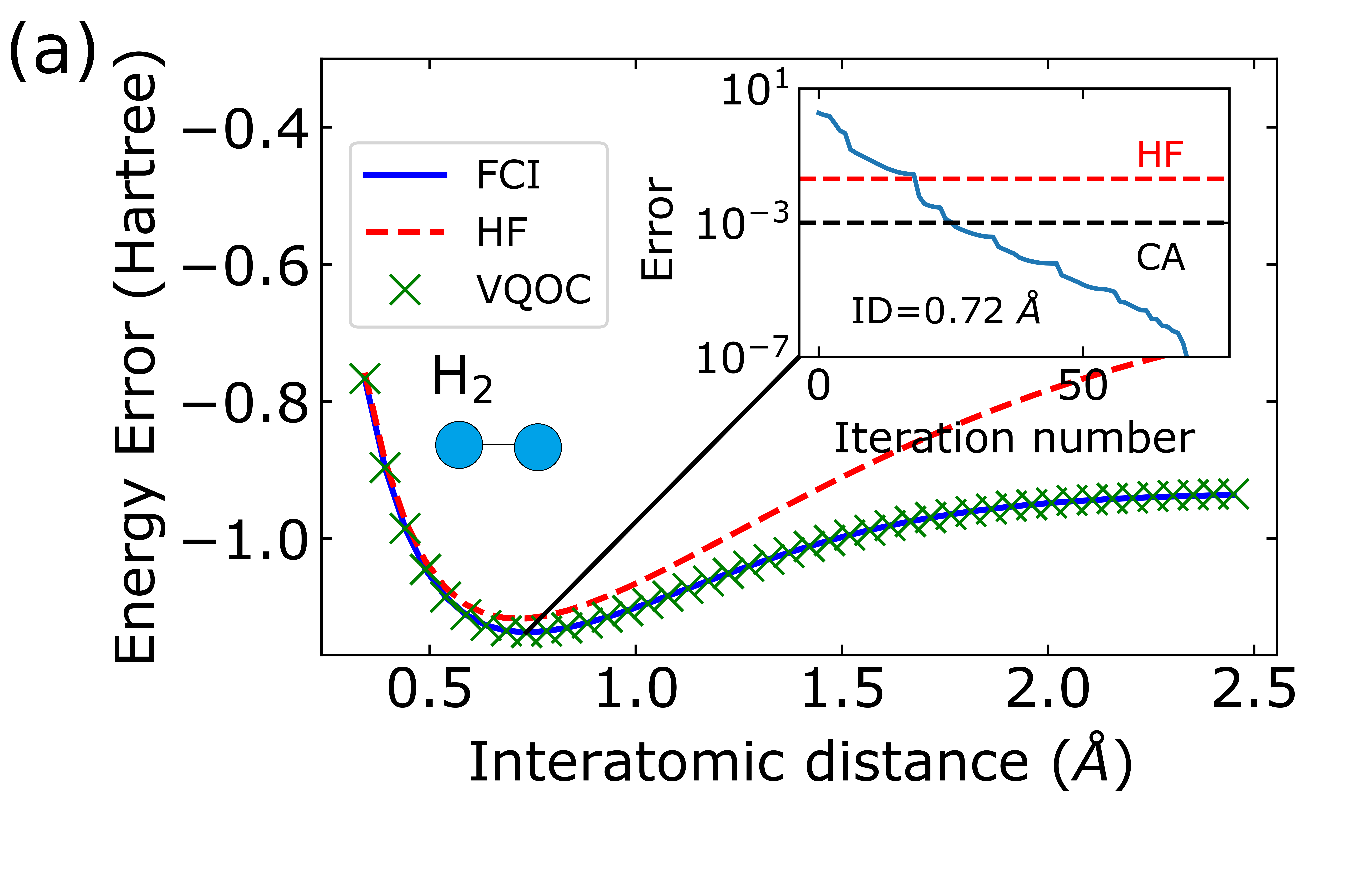}
    \end{minipage}%
    \begin{minipage}[t]{.49\textwidth}
         \includegraphics[width=\columnwidth]{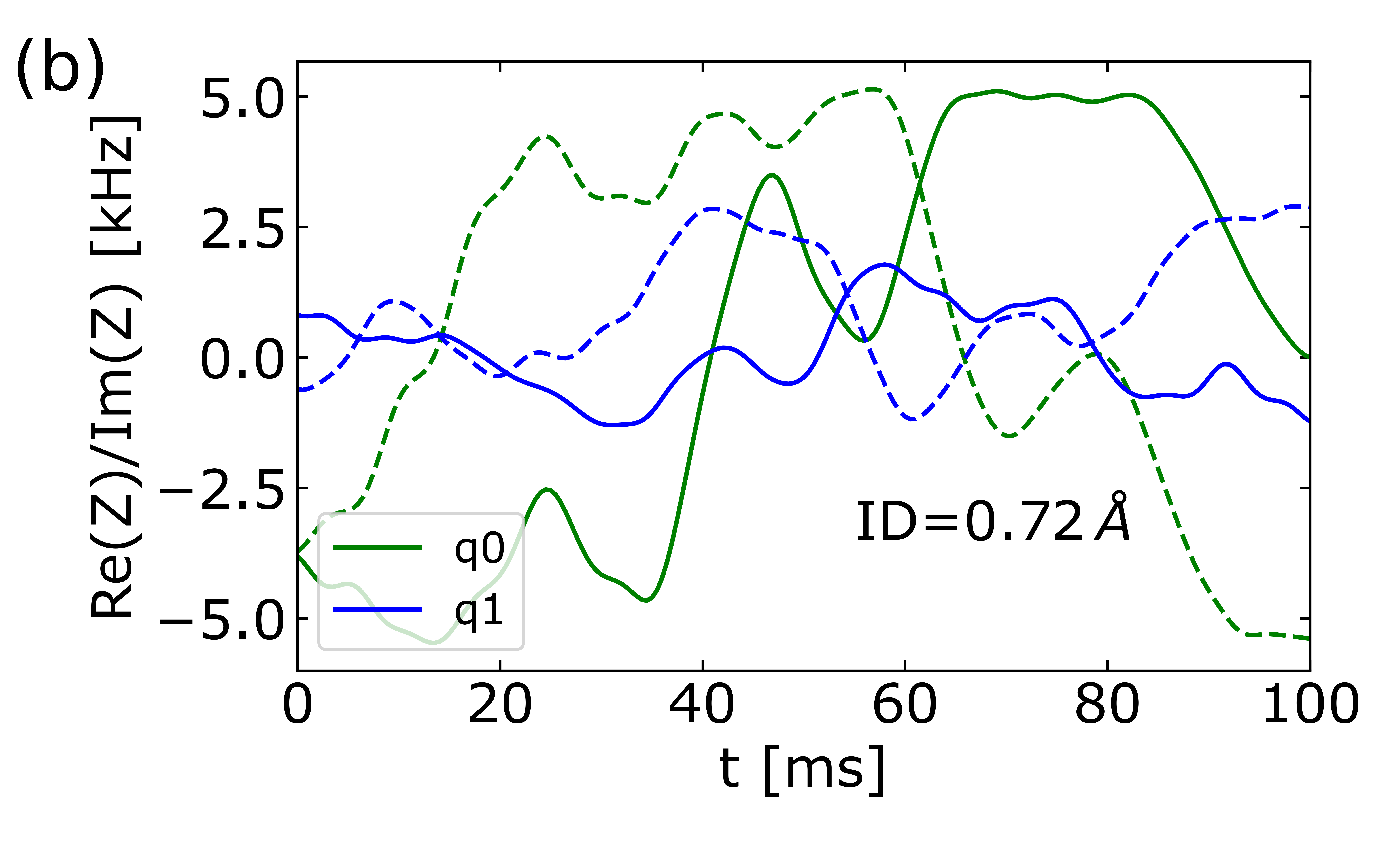}
    \end{minipage}%

\caption{Simulation results of the VQOC algorithm on H$_2$ with 2 qubits, Van der Waals drift Hamiltonian and rotational control. (a) Bond length energy potential landscape for H$_2$ with the FCI (exact), Hartree-Fock and VQOC energies. Inset: iteration vs.\ energy error w.r.t.\ FCI ground state energy at bond length equal to 0.72 \AA, together with Hartree-Fock energy and chemical accuracy. (b) Finalized pulses for process of (a) at ID=0.72\AA, showing real (full) and imaginary (dashed) parts of the rotation controls on the 2 qubits.}
    \label{fig:H2simulationsQOC}
    \end{centering}
\end{figure}
    \end{minipage}
\end{widetext2}

\begin{widetext2}
    \begin{minipage}[b]{\linewidth}
\begin{figure}[H]
\begin{centering}
\captionsetup{justification=raggedright}
    \begin{minipage}[t]{.49\textwidth}
         \includegraphics[width=\columnwidth]{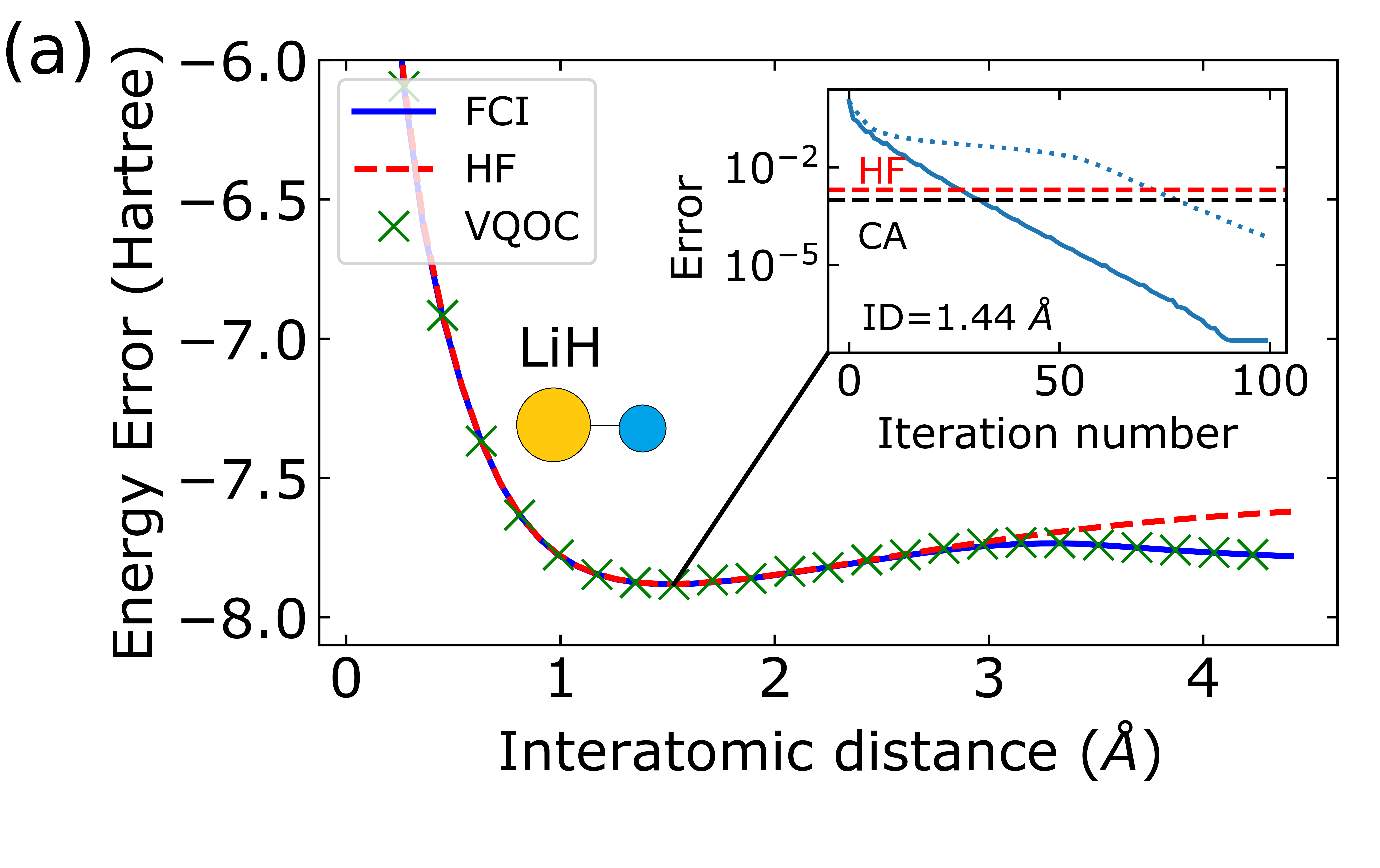}
    \end{minipage}%
    \begin{minipage}[t]{.49\textwidth}
         \includegraphics[width=\columnwidth]{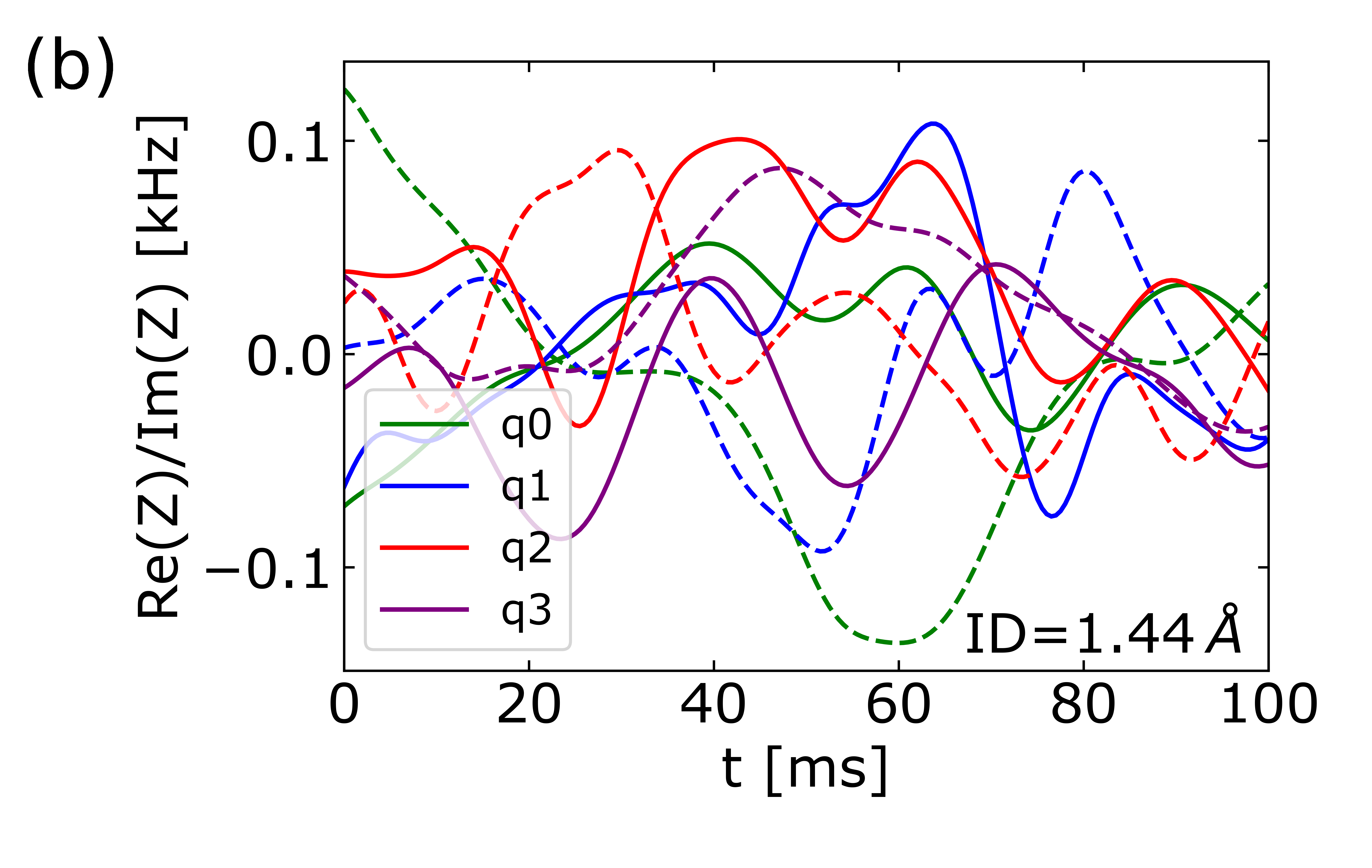}
    \end{minipage}%
\caption{Simulation results of the VQOC algorithm on LiH with 4 qubits, Van der Waals drift Hamiltonian and rotational control. (a) Bond length energy potential landscape for LiH, with the FCI (exact), Hartree-Fock and minimal VQOC energies. Inset: iteration vs.\ energy error w.r.t.\ FCI ground state energy at bond length equal to 1.44 \AA \ for minimal initial state (full) and mean over the initial states (dashed), together with Hartree-Fock energy and chemical accuracy. (b) Finalized pulses for process of (a) at ID=1.44 \AA, showing real (full) and imaginary (dashed) parts of the rotation controls on the 2 qubits for the best found result.}
    \label{fig:LIHsimulationsQOC}
    \end{centering}
\end{figure}
    \end{minipage}
\end{widetext2}

\begin{widetext2}
    \begin{minipage}[b]{\linewidth}
\begin{figure}[H]
    \centering
    \captionsetup{justification=raggedright}
    \includegraphics[width=0.78\textwidth]{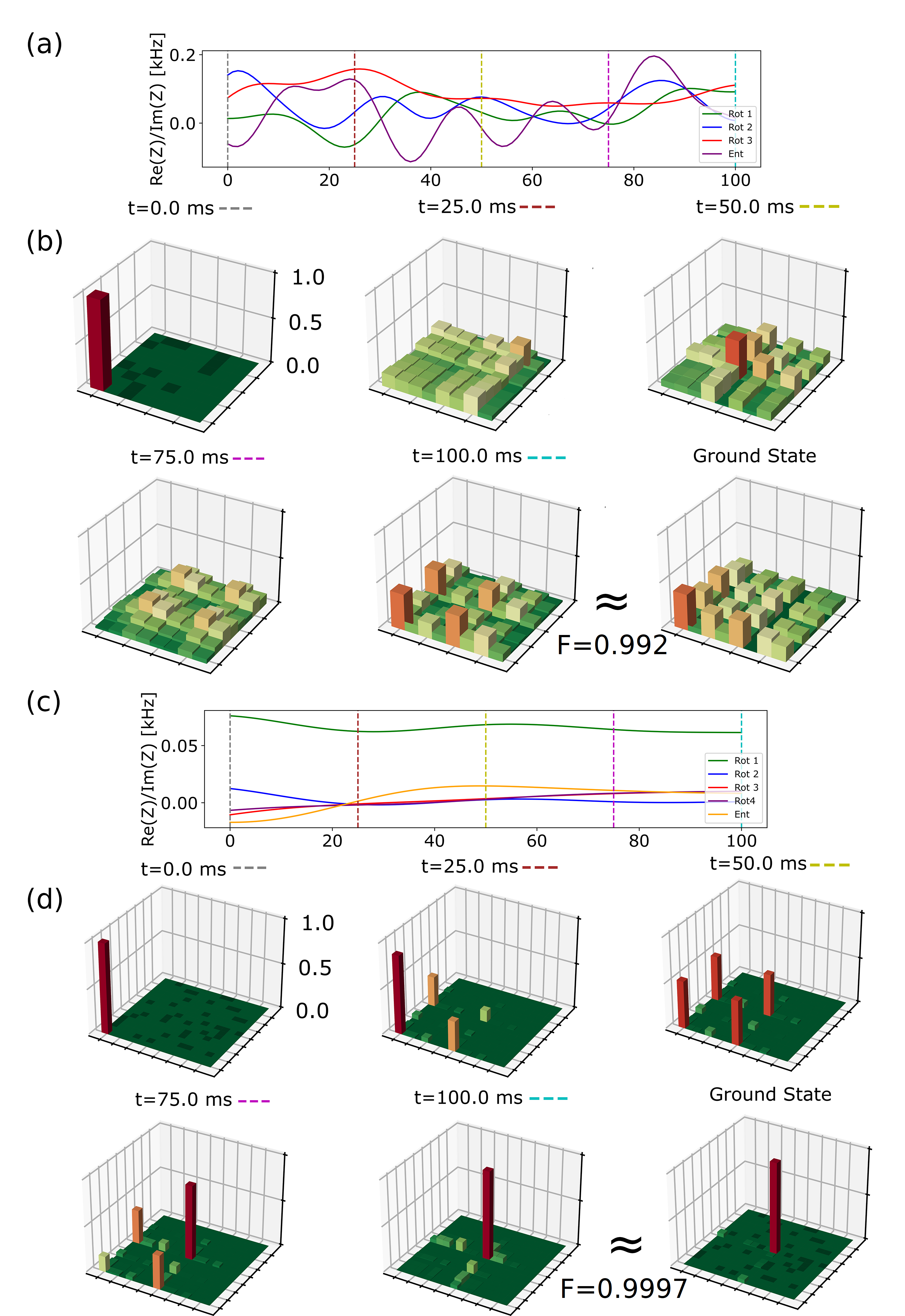}
    \caption{Ground state optimization of random $2^3\times2^3$ Hermitian matrix (3 qubits, from 30 random Pauli strings) and of 4-qubit LiH Hamiltonian at bond length equal to 0.99\AA, with Van der Waals drift Hamiltonian and rotational and entanglement control. (a) Finalized pulses showing real part of rotation and entanglement for random Hamiltonian. (b) Density matrix evolution plotted at different times during the evolution process together with the density plot of the ground state for random Hamiltonian and fidelity $F$. (c) Finalized pulses showing real part of rotation and entanglement for LiH Hamiltonian. (d) Density matrix evolution plotted at different times during the evolution process, together with the density plot of the ground state  for LiH Hamiltonian and fidelity $F$.}
    \label{fig:LIHsimulationsQOC2}
\end{figure}
    \end{minipage}
\end{widetext2}
\textcolor{white}{.}
\newpage
\textcolor{white}{.}
\newpage

To compare the VQE and VQOC algorithms as described in Sec.~\ref{sec:section5}, LiH and H$_4$ Hamiltonians are analyzed. Figures \ref{fig:VQEvsQOCLiH} and \ref{fig:VQEvsQOCH4}, respectively, show VQE vs.\ VQOC results for LiH with $d=9$ and H$_4$ with $d=4$, giving  $T_{\text{VQE}}=T_{\text{VQOC}}=99$ ms and $T_{\text{VQE}}=T_{\text{VQOC}}=44$ ms respectively. In both cases, all possible basis states are taken as initial states \cite{chemicalaccuracy} and the mean and best cases are reported. 
\medskip

The first thing to note is that for both types of entanglement, the best VQOC result generally outperforms the VQE results, while the average results are quite similar, even slightly better for VQE in the H$_4$ case. Moreover, the VQOC results are way below chemical accuracy, whereas this is not reached for VQE. For LiH in Fig.~\ref{fig:VQEvsQOCLiH}, the dipole interaction results perform well in the intermediate interatomic distances (where the ground state is highly entangled), while the Van der Waals interaction results perform better in the low and high interatomic distances (where the ground state is lowly entangled). For H$_4$, similar behaviour is seen in Fig.~\ref{fig:VQEvsQOCH4}, as VQOC performs better at low interatomic distances, where the ground state is lowly entangled, than at high interatomic stances where the ground state is highly entangled. The reason for this might have to do with the qubit architecture, e.g.\ for the LiH Hamiltonians, two non-neighbouring qubits (1 and 3) have to be entangled without being entangled to qubit 2, which is challenging given the strong nearest neighbour interactions in a Rydberg system. This indicates the importance of architecture choice in VQOC and will be a future point of investigation, as previously done in VQE \cite{entanglinggateinfluence}.

\begin{minipage}[b]{\linewidth}
\begin{figure}[H]
\begin{centering}
\captionsetup{justification=raggedright}
    \begin{minipage}[t]{0.99\textwidth}
         \includegraphics[width=1\columnwidth]{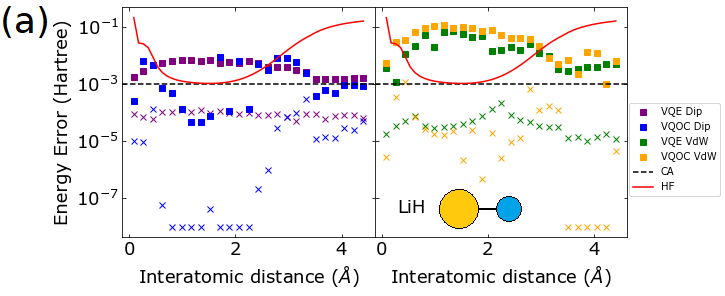}
    \end{minipage}%
    \\
    \begin{minipage}[t]{.99\textwidth}
         \includegraphics[width=1\columnwidth]{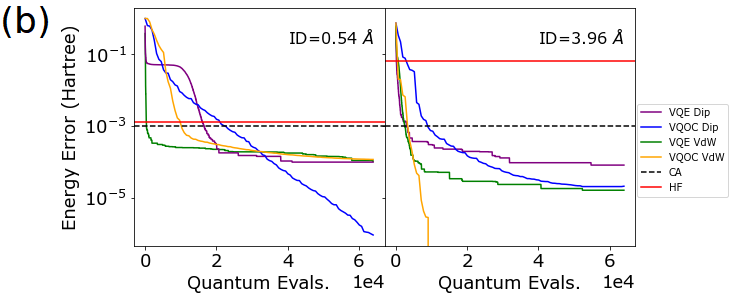}
    \end{minipage}%
\caption{Comparison of VQE vs.\ VQOC 
for LiH ($d_{\text{VQE}}=9\Rightarrow T_{\text{VQE}}=T_{\text{VQOC}}=99$ms). (a) Plot of final energy errors for different atomic distances after $6.4\cdot10^4$ total quantum evaluations. (b) Energy error vs.\ quantum evaluations for selected interatomic distances. VQE error is taken as the minimal energy achieved up until the specified quantum evaluation.}
    \label{fig:VQEvsQOCLiH}
    \end{centering}
\end{figure}
    \end{minipage}
    
Fig.~\hyperlink{fig:VQEvsQOC1b}{\ref{fig:VQEvsQOCLiH}b} shows two convergence behaviour plots. The left plot shows illustrative behaviour for the low interatomic regime where VQOC with dipole interaction performs best and the behaviour of the other methods is quite comparable. The right plot shows illustrative behaviour for the high interatomic regime, where VQOC is generally better than VQE for both interaction types, and the Van der Waals interaction excels. Fig.~\hyperlink{fig:VQEvsQOC1b}{\ref{fig:VQEvsQOCH4}b} shows convergence plots for H$_4$, where it is seen that VQOC is able to find energy levels below chemical accuracy, whereas VQE stifles.

\begin{minipage}[b]{\linewidth}
\begin{figure}[H]
\begin{centering}
\captionsetup{justification=raggedright}
    \begin{minipage}[t]{0.99\textwidth}
         \includegraphics[width=1\columnwidth]{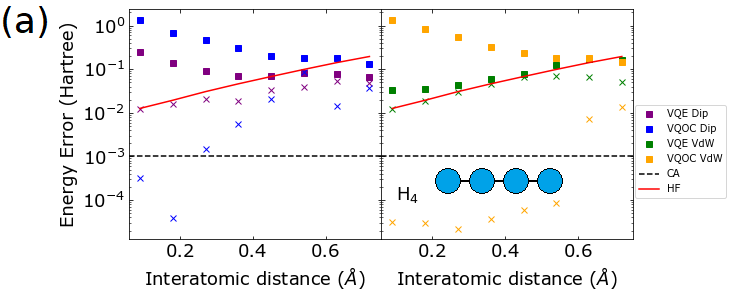}
    \end{minipage}%
    \\
    \begin{minipage}[t]{.99\textwidth}
         \includegraphics[width=1\columnwidth]{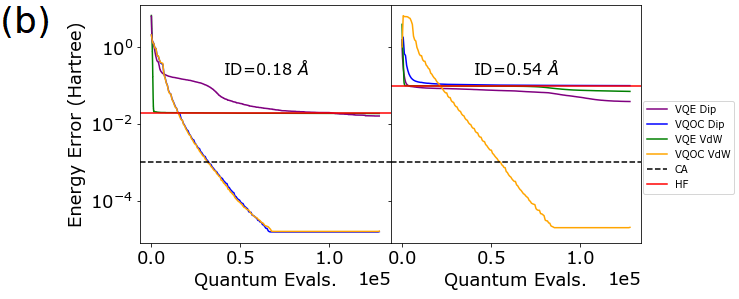}
    \end{minipage}%
\caption{Comparison of VQE vs.\ VQOC
 for H$_4$ 
($d_{\text{VQE}}=4\Rightarrow T_{\text{VQE}}=T_{\text{VQOC}}=44$ms). 
 (a) Plot of final energy errors for different atomic distances after $1.28\cdot10^5$ total quantum evaluations. (b) Energy error vs.\ quantum evaluations for selected interatomic distances. VQE error is taken as the minimal energy achieved up until the specified quantum evaluation.}
    \label{fig:VQEvsQOCH4}
    \end{centering}
\end{figure}
    \end{minipage}    

\subsection*{QSL Results}
To highlight the advantage of VQOC opening up a bigger search space at low evolution times $T$, we consider randomly generated 4-qubit Hamiltonians, generated as described at the beginning of this section. For VQE we again consider the hardware efficient ansatz with varying $d$, $\tau_g=1$ ms, Van der Waals entanglement gates with $\tau_V=10$ ms and entanglement time $T_{ent}\in[0,2\pi \tau_V]$ resulting in the entanglement gate $\exp(iH_dT_{ent})$. This results in total evolution times $T_{\text{VQE}}=d\cdot(\tau_g+T_{ent})$. We then consider the equivalent VQOC circuit, as described in Sec.~\ref{sec:comparing}, with $T_{\text{VQOC}}=T_{\text{VQE}}.$ Note that at time $T_{ent}=2\pi k \tau_V, k\in\mathbb{N}$, the Van der Waals entanglement gate is equal to the identity. In VQOC, something similar happens when $T_{\text{VQOC}}/N=2\pi k\tau_V, k\in\mathbb{N}$ as the step size becomes too small to capture the behaviour caused by the drift Hamiltonian.

\medskip

Fig.~\ref{fig:convergence_times}, shows an example of the errors w.r.t.\ the ground state of a randomly generated Hamiltonian after $64\cdot10^4$ total quantum evaluations. Here it can be seen that the reached errors and QSL are independent of circuit depth (for $d\geq 3$) for VQE and that VQOC reaches lower errors faster than VQE can. This supports the hypothesis that VQOC can open up a larger search space of states than VQE can, given a fixed evolution time. The errors increase again because of the entanglement gates transforming back to identity occurring at the expected times.

\begin{minipage}[b]{\linewidth}
\begin{figure}[H]
\begin{centering}
\captionsetup{justification=raggedright}
    \begin{minipage}[t]{.95\textwidth}
        \includegraphics[width=\columnwidth]{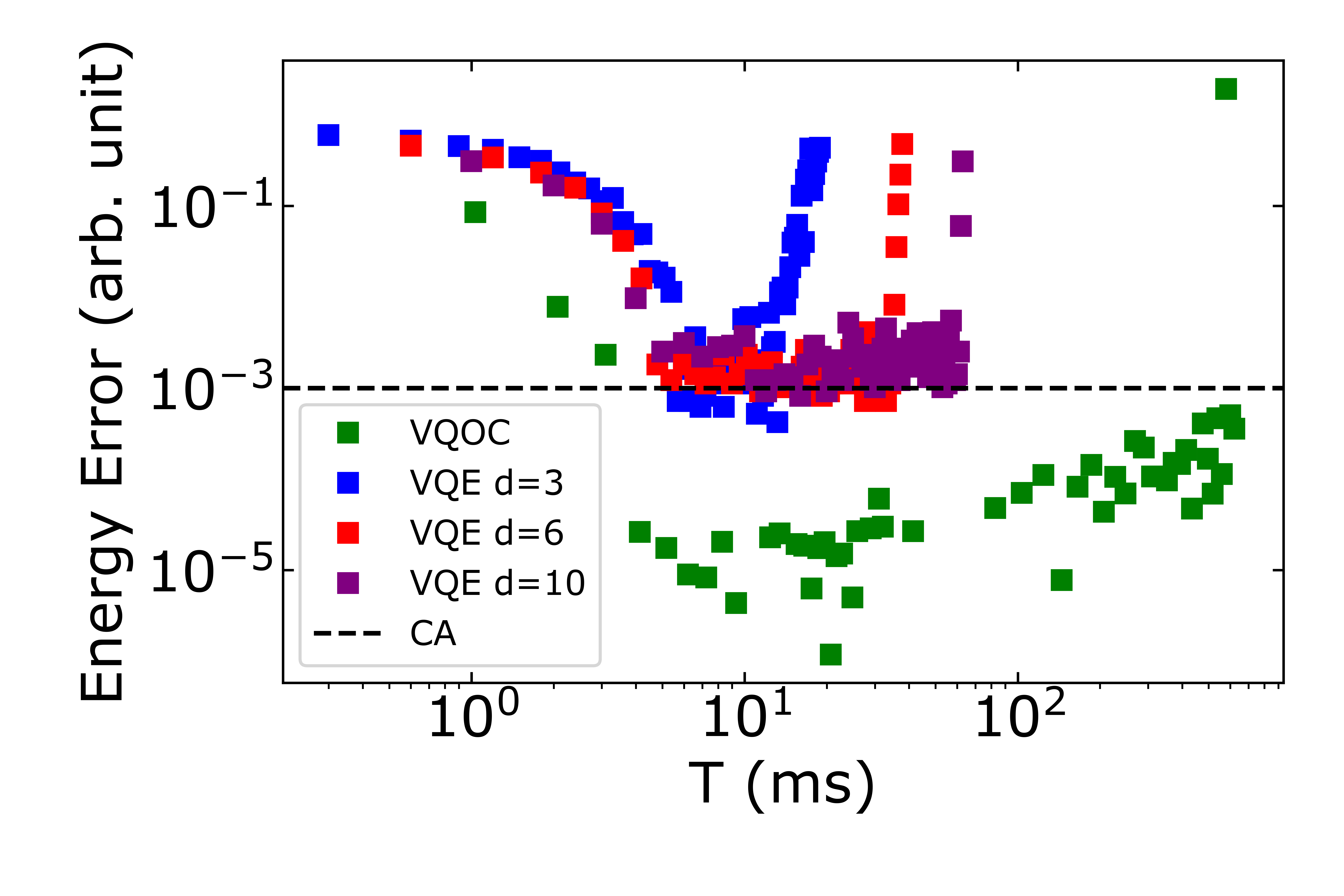}
    \end{minipage}%
    \caption{Errors achieved on a 4-qubit random Hamiltonian for VQOC and VQE at different values of $T_{\text{VQE}}=T_{\text{VQOC}}$ after $64\cdot 10^4$ total quantum evaluations.}
    \label{fig:convergence_times}
    \end{centering}
\end{figure}
\end{minipage}

\section{Conclusion}

In this work, we introduce a novel pulse-based variational quantum optimal control (VQOC) algorithm for quantum chemistry applications. The method has, through simulations, shown proof of concept for a neutral atom quantum computing system with interactions between the qubits mediated via Rydberg excitations, by approximating ground state energies of simple molecules with chemical accuracy. In this process, hyperparameters like the evolution time $T$ and the terms in the control Hamiltonian $H_c$ were found to play important roles.

\medskip

Comparing VQE and VQOC is not trivial, as both algorithms use entirely different optimization methods. For our criteria of quantum overhead, we have shown that our VQOC method can compete with, and in certain cases outperform VQE, while doing so in a more NISQ-friendly manner. Especially when considering equivalent evolution process at small depths (short $T$), VQOC can outperform VQE. In particular, the better convergence of VQOC in the low evolution time regime will help stifle the influence of decoherence in NISQ-era quantum computing problems. The reason for the faster convergence lies in the accessibility of the Hilbert space, and thus links to the QSL. In regimes with higher $T$, depending on the type of drift Hamiltonian introduced and the ground state of the problem Hamiltonian, VQOC either outperforms VQE or has similar convergence behaviour.

\medskip

The exact reason for this behaviour should be part of future research. However, we believe that for systems where the interatomic interactions are very strong, giving rise to a strongly-entangled ground state, VQOC is able to find this state faster due to entanglement being accounted for at every time step. Furthermore, there is a difference in convergence behaviour under varying entanglement schemes, which could be a result of a non-optimal qubit arrangement. An obvious solution to this would be to try and find optimal permutations of the qubits or fit the entanglement scheme to the problem Hamiltonian. A perhaps more interesting solution would be to treat the qubit positions as optimizable parameters (which due to technical constraints would be time-independent) in the VQOC scheme.

\medskip

Other points of future research would include the physical implementation of the algorithm. Specifically, the Rydberg version of the VQOC algorithm would be a perfect implementation on a neutral atom optical tweezer platform with Rydberg excitations. Several points to focus on in an experimental implementation are the repeatability of the algorithm, the robustness due to errors within the pulses and the required precision of the pulses, and creating specific controlled gates as required by the parameter gradient algorithm. Another implementation issue is the matter of noise. On the pulse-based level, certain noise terms can be introduced eloquently by switching to Lindbladian evolution instead of Hamiltonian \cite{noiselindblad}. A last interesting point would be to look into adapting the VQOC algorithm for qutrit or qudit manifolds \cite{quditpaper} and its influence on barren plateaus \cite{barrenplateau1,barrenplateau2}. 

\section*{Acknowledgements}
We thank Deon Janse van Rensburg, Jasper Postema and Madhav Mohan for discussions. This research is financially supported by the Dutch Ministry of Economic Affairs and Climate Policy (EZK), as part of the Quantum Delta NL programme, and by the Netherlands Organisation for Scientific Research (NWO) under Grant No.\ 680.92.18.05.

\section*{Data Availability}
The data that support the findings of this study are available from the corresponding author upon reasonable request.

\newpage
\bibliographystyle{apsrev4-1}
\bibliography{Bibliography}

\end{document}

\onecolumngrid
\section*{Supplementary Information: Pulse based Variational Quantum Optimal Control for hybrid quantum computing}

As mentioned in the main text, the goal in our optimal control setup is to minimize a given {\em cost functional} $J: \calU\times \calZ \mapsto \mathbb{R}$ subject to the equality constraint $e(U,z)=0$ for a given map $e: \calU\times \calZ\mapsto \calX^*$, where $\calZ=L^2((0,T);\mathbb{C}^L)$, $L\in\mathbb{N}$ with inner product \begin{equation*}
    \langle \tilde{z},z\rangle_\calZ:=\sum_{l=1}^L\int_{0}^T \text{Re}\left[\tilde{z}_l(t)\overline{z_l(t)}\right]\,dt,\qquad z,\tilde z\in \calZ
\end{equation*}
and $\calU=\calX = W^{1,2}((0,T);\calL(\calH^m))$ with the inner product
\begin{equation*}
    \langle \tilde P,P\rangle_{\calU} := \int_0^T \langle \partial_t\tilde P(t),\partial_t P(t)\rangle_F + \langle \tilde P(t),P(t)\rangle_F \,dt,\qquad P,\tilde P \in \calU.
\end{equation*}
Here, $e=0$ describes the {\em state equation}, which relates the unitary propagators $U\in\calU$ to control pulses $z\in\calZ_{ad}$, with $\calZ_{ad}$ as defined in the main text. More precisely, it describes the Schr\"odinger equation as a constraint in the following sense: $e(U,z)=0$ if and only if
\[
    \langle P,e(U,z)\rangle_{\calX,\calX^*} := \text{Re}\bigl[\langle P(0),U(0) - I\rangle_F\bigr] +\int_0^T \text{Re}\bigl[\langle P(t),i\partial_t U(t)-H[z(t)]U(t)\rangle_F\bigr]\,dt = 0\qquad\text{for all $P\in\calX$.}
\]
The existence of pairs $(U,z)$ satisfying $e(U,z)$ in our functional framework is discussed in Appendix~\ref{sec:wellposed}.

\medskip

The problem statement then reads
\begin{equation}
    \min_{(U,z)\in \calU\times \calZ_{ad}} J(U,z)\quad \text{subject to }\quad e(U,z)=0\;\;\text{in $\calX^*$},
    \label{eq:optimalcontrolform1}
\end{equation}
of which the solution is provided in Appendix~\ref{sec:minimizer}.

\medskip

To compute a minimizer of \ref{eq:optimalcontrolform1}, we will make use of the system of first-order optimality conditions, which we derive via the Lagrangian formulation of Eq.~\eqref{eq:optimalcontrolform1}. In this formulation, a Lagrange multiplier $P\in\calX$ is added to penalize the constraint $e=0$, i.e.\ we consider the Lagrangian $\mathscr{L}:\calU\times \calZ\times \calX\to \mathbb{R}\cup \{+\infty\}$ given by
\begin{equation}
 \mathscr{L}(U, z, P):=J(U, z)+\langle P, e(U, z)\rangle_{\calX, \calX^*}.
\label{eq:lagrangianformulation}
\end{equation}
The minimization problem \eqref{eq:optimalcontrolform1} then becomes the saddle-point problem
\begin{equation}
    \min_{(U,z)\in \calU\times\calZ_{ad}}\,\sup_{P\in \calX}\, \mathscr{L}(U, z, P).
\end{equation}

The necessary conditions for optimality can thus be reformulated as follows \cite{kkt}: 
\begin{equation}
    \begin{aligned}
D_P \mathscr{L}\bigl(U, z, P\bigr) &=0 & & \\%\text { in } \Omega, \\
D_U \mathscr{L}\bigl(U, z, P\bigr) &=0 & & \\%\text { in } \calX^{*}, \\
 D_z \mathscr{L}\bigl(U, z, P\bigr)[ \tilde z-z] & \geq 0 & & \enskip \forall \tilde z \in \calZ_{ad},
\end{aligned}
\label{eq:kktconditions}
\end{equation}
where $D_i\calL$ is the functional derivative of $\calL$ w.r.t.\ the parameter $i\in\{U,z,P\}$. The formal justification of the optimality condition is the subject of Appendix~\ref{app:frechetderivatives}.

\medskip

Finally, Appendix~\ref{app:gradient} discusses an approach in which one can evaluate gradients on a quantum machine.

\subsection{Well-posedness}
\label{sec:wellposed}
Here, we will prove the well-posedness of the Schr\"{o}dinger equation given by %\begin{equation}
\begin{equation}\label{eq:stateequation}
        i\partial_t U(t)=H[z(t)]U(t),\qquad
        U(0)=I, 
\end{equation}
for $z\in \calZ_{ad}$ and $U\in \calU$. This will also implicate well-posedness of the constraint in the main text
\begin{equation}
\begin{aligned}
    \langle P,e(U,z)\rangle_{\calX,\calX^*}=0\qquad\text{for all $P\in\calX$},
\end{aligned}
\end{equation}
which is the weak formulation of Eq.~\eqref{eq:stateequation}. To prove this, we define $\calU_\mathcal{C}:=\mathcal{C}_{b}([0,T],\mathcal{L}(\mathcal{H}^m))$, the Banach space of bounded continuous functions, with norm
\begin{equation}
    \|U\|_{\calU_\mathcal{C}}=\sup_{t\in[0,T]}\left(\|U(t)\|_F\,e^{-\lambda t}\right).
    \label{schrodingerappendix}
\end{equation}
Based on Eq. (7) of the main text, the Hamiltonian is taken to be of the form
\[
    H[z(t)] = H_d + H_c[z(t)],\qquad\text{with}\quad H_c[z(t)] := \sum_{l=1}^L \left[ Q_lz_l(t)+Q_l^\dagger\overline{z_l(t)}\right].
\]

We prove the following theorem

\begin{theorem}[Well-posedness of Schr\"{o}dinger equation]
For each $z\in \calZ_{ad}$, there exists a unique mild solution \normalfont{\cite{mildsolution}} $U_{z}$ to Eq.~\eqref{eq:stateequation} in $\calU_\mathcal{C}$, i.e.\ $U_z$ satisfies
\[
    U_z(t) = I - i\int_{0}^t H[z(s)]U(s)\,ds\qquad\text{for all $t\in[0,T]$}.
\]
This solution depends continuously on $z$. Furthermore, $U_{z}$ is unitary, and $U_{z}\in \calU$ with
\[
    \|U_z\|_\calU \le \eta(m,T,\|H_d\|_F,\|z\|_\calZ),
\]
where $\eta$ is an increasing locally bounded function.
\end{theorem}

\subsubsection*{\textbf{Existence and uniqueness}}
\label{sec:solexist}

For a fixed $z\in\calZ_{ad}$, we define the map $\Lambda_z:\calU_\mathcal{C}\mapsto \calU_\mathcal{C}$ given by
\begin{equation}
    \Lambda_z[U](t):=I-i\int_{0}^t H[z(s)]U(s)\,ds,
\end{equation}
which is the integrated formulation of Eq.~\eqref{eq:stateequation}. Now let $U$, $V\in \calU_\mathcal{C}$ and $W=U-V\in\calU_\mathcal{C}$ then
\begin{equation}
\begin{aligned}
    \left\|\Lambda_z[W](t)\right\|_F^2&=\left\|\int_{0}^tH[z(s)]W(s)\,ds\right\|_F^2\leq \left(\int_{0}^t\|H[z(s)]\|_F\|W(s)\|_F\,ds\right)^2\\
    &=\left(\int_{0}^t\|H[z(s)]\|_F\|W(s)\|_F\,e^{-\lambda s}e^{\lambda s}\,ds\right)^2\leq\left( \sup_{s\in(0,T)}\|W(s)\|_F\,e^{-\lambda s}\right)\left(\int_{0}^t\|H[z(s)]\|_F\,e^{\lambda s}\,ds\right)\\
    &\leq \frac{c_H\|W\|_{\calU_\mathcal{C}}}{\lambda}(e^{\lambda t}-1),
\end{aligned}
\end{equation}
where $c_H=\esssup_{t\in(0,T)}\|H[z(t)]\|_F$, which exists since $z\in\calZ_{ad}$. The following inequality follows
\begin{equation}
    \|\Lambda_z[W](t)\|_F^2\,e^{-\lambda t}\leq \frac{c_H\|W\|_{\calU_\mathcal{C}}}{\lambda}(1-e^{-\lambda t})\leq \frac{c_H\|W\|_{\calU_\mathcal{C}}}{\lambda}\quad\Longleftrightarrow\quad \|\Lambda_z[W]\|_{\calU_\mathcal{C}}\leq\frac{c_H\|W\|_{\calU_\mathcal{C}}}{\lambda}.
\end{equation}
Taking $\lambda$ large enough implies that $\Lambda_z$ is a contraction in the $\calU_\mathcal{C}$-norm. Since $\calU_\mathcal{C}$ is complete, the Banach fixed-point theorem can be invoked to find a fixed point $U_z$ of $\Lambda_z$ so that 
\begin{equation}
\begin{aligned}
    U_{z}(t)=\Lambda_z[U_{z}](t)=I-i\int_{0}^t H[z(s)]U_{z}(s)\,ds. 
    \label{eq:appa7}
\end{aligned}
\end{equation}
This proves both existence and uniqueness of a mild solution $U_{z}\in\calU_\mathcal{C}$. We in fact have more. By virtue of $U_z\in \calU_\mathcal{C}$ and $H[z]\in L^\infty((0,T);\calL(\calH^m))$, the product $H[z]U_z\in L^2((0,T);\calL(\calH^m))$. Hence, $t\mapsto U_z(t)$ is absolutely continuous with weak derivative $\partial_t U_z \in L^2((0,T);\calL(\calH^m))$ due to Eq.~\eqref{eq:appa7}. In particular, $U_z\in\calU$.

\medskip

The unitarity of $U_{z}$ is shown using the product rule
\begin{equation}
\begin{aligned}
    \frac{d}{dt}\Bigl[U_{z}^\dagger(t)U_{z}(t)\bigr] 
    &=iU_z^\dagger(t) H[z(t)]^\dagger U_z(t)-iU_z^\dagger(t) H[z(t)]^\dagger U_z(t)=0.
    \end{aligned}
\end{equation}
Thus, $U_{z}^\dagger(t)U_{z}(t)=U_0^\dagger U_0=I$. The proof for $U_{z}(t)U_{z}^\dagger(t)=I$ follows analogously. As for the bound, we notice that
\begin{equation}
    \|U_{z}\|_{L^2((0,T);\mathcal{L}(\mathcal{H}^m))}^2=\int_{0}^T \|U_{z}(t)\|_F^2\,dt=2^mT.
\end{equation}
Moreover, the submultiplicativity of the Frobenius norm yields
\begin{equation}
\begin{aligned}
       \|\partial_tU_{z}\|&_{L^2((0,T);\mathcal{L}(\mathcal{H}^m))}^2=\int_{0}^T \|\partial_tU_{z}(t)\|_F^2\,dt\le \int_{0}^T \|H[z(t)]\|_F^2\|U_{z}(t)\|_F^2\,dt\\
       &=\int_{0}^T 2^m\|H_d+H_c[z(t)]\|_F^2\,dt\leq 2^{m+1}\int_{0}^T \|H_d\|^2_F+\|H_c[z(t)]\|_F^2\,dt\\
       &\leq2^{m+1}T\|H_d\|^2_F+2^{m+1}\int_{0}^T \left\|\sum_{l=1}^Lz_l(t)Q_l+\overline{z}_l(t)Q_l^\dagger\right\|_F^2 dt\\
       &\leq2^{m+1}T\|H_d\|^2_F+2^{m+3}Q_{\text{max}}^2\int_{0}^T \sum_{l=1}^L|z_l(t)|^2\,dt\leq2^{m+1}T\|H_d\|^2_F+2^{m+3}Q_{\text{max}}^2\|z\|^2_{\calZ}.\\
\end{aligned}
\end{equation}
where $Q_{max}:=\max_{l=1,...,L}\|Q_l\|_F$. In total
\begin{equation}
\begin{aligned}
    \|U_{z}\|_{\calU}^2\leq 2^m \left(T+2T\|H_d\|_F^2+8Q_{\text{max}}^2\|z\|_{\calZ}^2\right) =: [\eta(m,T,\|H_d\|,\|z\|_\calZ)]^2.
    \label{eq:sobolevestimate}
\end{aligned}
\end{equation}
In conclusion, for each $z\in \calZ_{ad}$ a unique operator $U_z\in \calU$ exists as the solution of $U=\Lambda_z[U]$. A bounded solution operator $S:\calZ_{ad}\to \calU$ can be defined as $S(z)=U_z$ where $U_z=\Lambda_z[U_z]$.

\medskip

Now since $U_z(0)=I$ and $U_z\in \calU$ we have that $\langle P,e(U_z,z)\rangle_{\calX,\calX^*}=0$ for all $p\in\calX$.

\subsubsection*{\textbf{Continuity w.r.t. initial data}}

We show below that the solution operator $S:\calZ_{ad}\to \calU$ defined above is Lipschitz continuous, form which continuity follows.

\medskip

Let $z$, $w\in\calZ_{ad}$, and consider their corresponding solutions $U_z$, $U_w\in \calU$ respectively. Writing down the equation for the difference between $U_z$ and $U_w$ yields
\begin{equation*}
U_z(t)-U_w(t)=-i\int_0^t \bigl[H_d+H_c[z(s)]\bigr]\bigl[U_z(s)-U_w(s)\bigr]-\bigl[H_c[z(s)]-H_c[w(s)]\bigr]U_w(s)\,ds
\end{equation*}
Taking the Frobenius norm and using the triangle inequality gives
\begin{equation}
    \begin{aligned}
            \|U_z(t)-U_w(t)\|_F&=\int_0^t \|H_d+H_c[z(s)]\|_F\|U_z(s)-U_w(s)\|_F+\|H_c[z(s)]-H_c[w(s)]\|_F\|U_w(s)\|_F\,ds\\
            &\leq\int_0^t H_{\text{max}}\|U_z(s)-U_w(s)\|_F+\|H_c[z(s)]-H_c[w(s)]\|_F\,ds,
            \label{eq:appa16}
    \end{aligned}
\end{equation}
where $\|U_w(s)\|_F=1$ for every $s\in[0,T]$ by unitariness and $H_{\text{max}}:=\sup_{t\in[0,T]}\|H_d+H_c[z(t)]\|_F$ exists since $z\in \calZ_{ad}$, which is bounded. We then estimate the term
\begin{equation}
\begin{aligned}
    \|H_c[z(s)]-H_c[w(s)]\|_F&=\left\|\sum_{l=1}^L\bigl(z_l(s)-w_l(s)\bigr)Q_l+\bigl(\,\overline{z_l(s)}-\overline{w_l(s)}\,\bigr)Q_l^\dagger\right\|_F\\
    &\leq 2Q_{\text{max}}\sum_{l=1}^L|z_l(s)-w_l(s)|,
\end{aligned}
\end{equation}
where again $Q_{max}:=\max_{l=1,...,L}\|Q_l\|_F$. Using this result in Eq.~\eqref{eq:appa16}, we obtain
\begin{equation}
\begin{aligned}
    \|U_z(t)-U_w(t)\|_F&\leq\int_0^t H_{\text{max}}\|U_z(s)-U_w(s)\|_F+2Q_{\text{max}}\sum_{l=1}^L|z_l(s)-w_l(s)|\,ds\\
    &\leq 2\sqrt{T}Q_{\text{max}}\|z-w\|_\calZ+\int_0^t H_{\text{max}}\|U_z(s)-U_w(s)\|_F\,ds.
\end{aligned}
\end{equation}
Gr\"{o}nwalls Lemma \cite{gronwall} now implies
\begin{equation}
\|U_z(t)-U_w(t)\|_F
\leq 2\sqrt{T}Q_{\text{max}}\, e^{H_{\text{max}} T}\|z-w\|_{\calZ}.
\end{equation}
Since the RHS is independent of $t$
\begin{equation}
    \|U_z-U_w\|_{\calU_\mathcal{C}}\leq\sup_{t\in(0,T)}\|U_z(t)-U_w(t)\|_F\leq  2\sqrt{T}Q_{\text{max}}\, e^{H_{\text{max}}T}\|z-w\|_{\calZ},
\end{equation}
from which the Lipschitz continuity follows. This together with the existence and uniqueness proves the well-posedness of the Schr\"{o}dinger equation considered. 

\subsection{Minimizer Existence}\label{sec:minimizer}

One of the main results of optimal control theory is the existence of minimizers for problems such as Eq.~\eqref{eq:optimalcontrolform1} \cite{optimalcontrol}. The following theorem is an adaptation of existence theorems for optimization problems with differential equations as constraints, whose proof is included for completeness.

\begin{theorem}[Existence Minimizer]
The functional $J$ as in Eq.~\eqref{eq:optimalcontrolform1} attains a global minimum on $\calU\times \calZ_{ad}$ under the following assumptions:

\begin{enumerate}[label=(A\arabic*)]
\item $\calZ_{ad}\subset \calZ$ is a weakly closed subset of a Hilbert space;
\label{enumerationA1}
\item The state space $\calX$ is a Hilbert space;
\label{enumerationA2}
\item The state equation has a bounded solution operator $S:\calZ_{ad}\mapsto \calU$, i.e. \label{enumerationA3}
\[ 
\|S(z)\|_\calU\leq M(\|z\|_\calZ),\qquad\text{where\; $M:[0,\infty)\mapsto[0,\infty)$ is a locally bounded function;}
\]
\item The map $(U,z)\in \calU\times \calZ_{ad} \mapsto  e(U,z)\in \calX^*$ is weakly continuous;
\label{enumerationA4}
\item $J$ is weakly lower semicontinuous;
\label{enumerationA5}
\item $\calZ_{ad}$ is bounded.
\label{enumerationA6}
\end{enumerate}
\label{thm:existence}
\end{theorem}

\begin{proof}
To prove the theorem, $f:\calZ_{ad}\mapsto \mathbb{R}$ is defined as $f(z):=J(S(z),z)$ (using \ref{enumerationA3}) and shown to be weakly lower semicontinuous. Let $\{z^{(k)}\}_{k\ge 1}\subset \calZ_{ad}$ be a sequence such that $z^{(k)}\rightharpoonup \tilde{z}$ in $\calZ$. Then $\tilde{z} \in \calZ_{ad}$ by \ref{enumerationA1}. The $z^{(k)}$ are uniformly bounded in $\calZ$ \ref{enumerationA6}. Thus, there exists a $z_{\max}$ such that $\|z^{(k)}\|_{\calZ}\leq z_{\max}$. Consequently, $\|U_{z_l}\|_{\calU}\leq U_{max}$ by \ref{enumerationA3}. Reflexivity of the Hilbert space $\calU$ \ref{enumerationA2} then implies the existence of a subsequence and a limit point $\tilde U\in\calU$ such that $U_{z^{(k_l)}}\rightharpoonup\tilde{U}$ in $\calU$ \cite{evanspde}.

\medskip

From the weak continuity of $e:\calU\times \calZ_{ad}\mapsto \calX^*$ \ref{enumerationA4} and the weak lower semicontinuity of norms follows that
\begin{equation}
   \langle P,e(\tilde{U},\tilde{z})\rangle_{\calX,\calX^*}= \lim_{l\rightarrow\infty}\langle P, e(U_{z^{(k_l)}},z^{(k_l)})\rangle_{\calX,\calX^*}=0\quad \forall P\in \calX.
\end{equation}
Thus, $e(\tilde{U},\tilde{z})=0$, i.e.\ $\tilde{U}=U_{\tilde{z}}$. Together with the weak lower semicontinuity of $J$ \ref{enumerationA5}, this proves weak lower semicontinuity of $f$.

\medskip

Let $\{z^{(k)}\}_{k\ge 1}$ be a minimizing sequence, i.e.\  $\lim_{k\rightarrow\infty} f(z^{(k)})=\inf_{z\in \calZ_{ad}}f(z)$. Since $\calZ_{ad}$ is bounded \ref{enumerationA6} there must exist a $\tilde{z}$ and a subsequence $\{z^{(k_l)}\}_{l=1}^\infty$ such that $z^{(k_l)}\rightharpoonup \tilde{z}$ in $\calZ.$ Moreover, $\tilde{z}\in \calZ_{ad}$ because $\calZ_{ad}$ is weakly closed \ref{enumerationA1}. Weak lower semicontinuity of $f$ then implies
\begin{equation}
    f(\tilde{z})\leq \liminf_{l\rightarrow \infty} f(z^{(k_l)})=\inf_{z\in \calZ_{ad}}f(z).
\end{equation}
This proves existence of a $\tilde{z}\in \calZ_{ad}$ and $\tilde{U}=S(\tilde{z})\in \calU$ achieving minimization of $J$ with $e(\tilde{U},\tilde{z})=0$.
\end{proof}

\medskip

This rest of this section proves the existence of global minimizers in the VQOC setting.

\begin{theorem}[Existence Minimizer VQOC]
Let $\calZ=L^2((0,T);\mathbb{C}^L)$ and $\calZ_{ad}=\{z\in\calZ \,|\, \|z\|_\infty\leq z_{\max}\}$. Furthermore, let $\calU=\calX=W^{1,2}((0,T);\calL(\calH^m))$. For any $|\psi_0\rangle\in\calH^m$, the minimization problem given by
\begin{equation}
    \min_{U\in \calU, z\in \calZ_{ad}} \langle\psi(T)|H_{\text{mol}}|\psi(T)\rangle+\frac{\lambda}{2}\langle z,z\rangle_\calZ=:J_1(U)+J_2(z),\quad |\psi(T)\rangle=U(T)|\psi_0\rangle
    \label{eq:costfunctional2}
\end{equation}
under the duality pairing constraint
\begin{equation}
\begin{aligned}
    \forall P\in\calX:\quad\langle P,e(U,z)\rangle_{\calX,\calX^*}:= \langle P(0),U(0) - I\rangle_F +\int_0^T \langle P(t),i\partial_t U(t)-H[z(t)]U(t)\rangle_F\,dt=0,
\end{aligned}
\end{equation}
attains a global minimum on $\calU \times \calZ_{ad}.$
\end{theorem}

\begin{proof}
To prove the theorem it is shown that the system adheres to \ref{enumerationA1}-\ref{enumerationA6} after which Theorem~\ref{thm:existence} is applied. Adherence to \ref{enumerationA1} follows from the weak lower semicontinuity of $\|\cdot\|_\infty$ and \ref{enumerationA6} is satisfied since $\|z\|_\calZ\leq C\|z\|_\infty$ for some $C>0$. \ref{enumerationA2} follows from the fact that $\calU$ is a Hilbert space. In Sec.~\ref{sec:wellposed} the existence of a locally bounded solution operator $S:\calZ_{ad}\mapsto \calU$ \ref{enumerationA3} is shown. $e: \calU\times \calZ_{ad}\mapsto \calX^*$ being weakly continuous \ref{enumerationA4} is shown by Lemma~\ref{app:a4} below, and the weak lower semicontinuity of $J$ follows trivially from the continuity of $J$ \ref{enumerationA5}. All conditions are adhered to and a minimum can be achieved.
\end{proof}

\begin{lemma}[Condition (A4)]
\label{app:a4}
Let $e:\calU\times \calZ_{ad}\mapsto \calX^*$ as in Eq.~\eqref{eq:stateequation}. Then $e$ is weakly closed, i.e.
\begin{equation}
    (U^{(k)},z^{(k)})\rightharpoonup (U,z)\quad\text{in $\calU\times \calZ$}\quad \Longrightarrow\quad e(U^{(k)},z^{(k)})\rightharpoonup e(U,z)\quad\text{in $\calX^*$}.
\end{equation}
\end{lemma}

\begin{proof}
This result is proven by proving that weak convergence of $z^{(k)}\rightharpoonup z$ in $\calZ$ implies weak convergence of $H[z^{(k)}]\rightharpoonup H[z]$ in $L^2((0,T);\mathcal{L}(\mathcal{H}^m))$. Then it is shown that weak convergence of $U^{(k)}\rightharpoonup U$ in $\calU$ will imply strong convergence of $\{U^{(k)}\}_{k\ge 1}$ in appropriate spaces. These facts together will imply that $e(U^{(k)},z^{(k)})\rightharpoonup e( U, z)$ in $\calX^*.$

\medskip

The pulses (control functions) are modelled as complex functions $z\in \calZ_{ad}$. Assume $z^{(k)}\rightharpoonup  z$ in $\calZ$. Then,
\begin{equation}
    \begin{aligned}
    \langle H[z^{(k)}],A\rangle_{L^2((0,T);\mathcal{L}(\mathcal{H}^m))}&=
    \int_{0}^T \left\langle\left(H_d+\sum_{l=1}^L z^{(k)}_{l}(t)Q_l+\overline{z^{(k)}_{l}(t)}Q_l^\dagger\right),\, A(t)\right\rangle_F\,dt\\
    &=\int_{0}^T\langle H_d, A(t)\rangle_F\,dt +\int_{0}^T \sum_{l=1}^L  z^{(k)}_{l}(t)\langle Q_l, A(t)\rangle_F\,dt+\sum_{l=1}^L  \overline{z^{(k)}_{l}(t)}\langle Q_l^\dagger, A(t)\rangle_F\,dt\\
    &\longrightarrow   \int_{0}^T\langle H_d, A(t)\rangle_F\,dt +\int_{0}^T \sum_{l=1}^L   z_{l}(t)\langle Q_l, A(t)\rangle_F\,dt+\sum_{l=1}^L  \overline{ z_{l}(t)}\langle Q_l^\dagger, A(t)\rangle_F\,dt\\
    &=\int_{0}^T \left\langle\left(H_d+\sum_{l=1}^L  z_{l}(t)Q_l+\overline{ z_{l}(t)}Q_l^\dagger\right),\, A(t)\right\rangle_F\,dt=\langle H[ z], A\rangle_{L^2((0,T);\calL(\calH^m))},\\
    \end{aligned}
\end{equation}
for any $A\in L^2((0,T);\mathcal{L}(\mathcal{H}^m))$, thus implying the weak convergence $H[z^{(k)}]\rightharpoonup H[ z]$ in $L^2((0,T);\mathcal{L}(\mathcal{H}^m))$.

\medskip

To eventually show weak continuity of the map $e$ \ref{enumerationA4} requires that every bounded sequence in $\calU_{ad}:=S(\calZ_{ad})$ has a strongly convergent subsequence (with limit not necessarily in $\calU_{ad}$). One way to prove this is by using 
the Rellich--Kondrachov embedding theorem \cite{evanspde}.

\begin{theorem}[Rellich--Kondrachov]
For any $n\ge 1$:
\begin{equation}
W^{1, 2}((0,T);\mathbb{R}^n) \hookrightarrow\hookrightarrow  C^{0}([0,T];\mathbb{R}^n).
\end{equation}

\end{theorem}

In Section~\ref{sec:wellposed}, it is shown that $\calU_{ad}\subset \calU= W^{1,2}((0,T);\calL(\calH^m))$. By noting that $\mathbb{C}^n\cong \mathbb{R}^{2n}$, the Rellich--Kondrachov Theorem can be applied. Thus, the compact embedding implies that every bounded sequence $\{U^{(k)}\}_{k\ge 1}
\subset \calU_{ad}$ 
has a subsequence $\{U^{(k_\ell)}\}_{\ell\ge 1}$ and some $\tilde U\in C^{0}([0,T];\mathcal{L}(\mathcal{H}^m))$ such that $\lim_{\ell\to\infty}\|U^{(k_\ell)} - \tilde U\|_{\sup}=0$. Since the full sequence weakly converges to $U$, and weak limits are unique, we deduce that $\tilde U$ is a continuous version of $U$ with $\tilde U=U$ almost everywhere. Furthermore, any strongly converging subsequence must convergence to $\tilde U$, thereby asserting that the full sequence $\{U^{(k)}\}_{k\ge 1}$ converges strongly in $C^{0}([0,T];\mathcal{L}(\mathcal{H}^m))$ towards $\tilde U$ \cite{kreyszig}.

\medskip

With the knowledge that $H[z^{(k)}]\rightharpoonup H[z]$ in $L^2((0,T);\mathcal{L}(\mathcal{H}^m))$ and $U^{(k)}\rightarrow\tilde{U}$ in $C^{0}([0,T];\mathcal{L}(\mathcal{H}^m))$, weak continuity of $e:\calU\times\calZ_{ad}\mapsto\calX^*$ can be proven. Indeed, let $P\in \calX$. Then,
\begin{equation}
\begin{aligned}
    &\langle P, e(U^{(k)},z^{(k)})\rangle_{\calX\times\calX^*}-\langle P, e(\tilde{U},z)\rangle_{\calX \times \calX^*}\\
    & =\underbrace{\text{Re}\left(\int_{0}^T \langle P(t), \partial_t U^{(k)}(t)-iH[z^{(k)}(t)]U^{(k)}(t)\rangle_F\,dt\right)-\text{Re}\left(\int_{0}^T \langle P(t), \partial_t \tilde{U}(t)-iH[z(t)]\tilde{U}(t))\rangle_F\,dt\right)}_{=:\text{(I)}}\\
    &+\underbrace{\text{Re}\bigl[\langle P(0),(U^{(k)}(0)-I)\rangle_F\bigr]-\text{Re}\bigl[\langle P(0),(\tilde{U}(0)-I)\rangle_F\bigr]}_{=:\text{(II)}}
\end{aligned}
\end{equation}
For (I), we have
\begin{equation}
    \begin{aligned}
    &\text{Re}\left(\int_{0}^T \langle P(t), \partial_t U^{(k)}(t)-iH[z^{(k)}(t)]U^{(k)}(t)\rangle_F\,dt\right)-\text{Re}\left(\int_{0}^T \langle P(t), \partial_t \tilde{U}(t)-iH[z(t)]\tilde{U}(t)\rangle_F\,dt\right)\\
    &=\text{Re}\left(\int_{0}^T \langle P(t), \partial_t\bigl( U^{(k)}(t)- \tilde{U}(t)\bigr)\rangle_F\,dt\right)-\text{Re}\left(i\int_{0}^T\langle P(t),H[z^{(k)}(t)](U^{(k)}(t)-\tilde{U}(t))\rangle_F\,dt\right)\\
    &-\text{Re}\left(i\int_{0}^T \langle P(t), \bigl(H[z^{(k)}(t)]-H[z(t)]\bigr)\tilde{U}(t)\rangle_F\,dt\right).
\end{aligned}
\end{equation}
The first term here goes to zero since $U^{(k)}\rightharpoonup\tilde{U}$ in $\calU$ so that $\partial_t U^{(k)}\rightharpoonup \partial_t \tilde{U}$ in $L^2((0,T);\mathcal{L}(\mathcal{H}^m))$. The third term goes to zero since we can write 
\begin{equation}
    \text{Re}\left(i\int_{0}^T \left\langle P(t), \bigl[H[z^{(k)}(t)]-H[z(t)]\bigr]\tilde{U}(t)\right\rangle_F\,dt\right)=    \text{Re}\left(i\int_{0}^T \left\langle P(t)\tilde{U}^\dagger(t), H[z^{(k)}(t)]-H[z(t)]\right\rangle_F\,dt\right),
\end{equation}
where now $P\tilde{U}^\dagger\in L^2((0,T);\mathcal{L}(\mathcal{H}^m))$ since $\tilde{U}^\dagger\in C^0([0,T];\mathcal{L}(\mathcal{H}^m))$ by Rellich--Kondrachov. As for the second term,
\begin{equation}
\begin{aligned}
    &\left|\text{Re}\left(i\int_{0}^T\left\langle P(t),H[z^{(k)}(t)]\bigl(U^{(k)}(t)-\tilde{U}(t)\bigr)\right\rangle_F\,dt\right)\right|\leq \int_{0}^T\left|\left\langle P(t),H[z^{(k)}(t)](U^{(k)}(t)-\tilde{U}(t))\right\rangle_F\right|\,dt\\ &\leq\int_{0}^T\|P(t)\|_F\bigl\|H[z^{(k)}(t)]\bigr\|_F\|U^{(k)}(t)-\tilde{U}(t)\|_F\,dt\\
    &\leq \esssup_{t\in(0,T)}\bigl\|H[z^{(k)}(t)]\bigr\|_F \|P\|_{L^2((0,T);\mathcal{L}(\mathcal{H}^m))}\|U^{(k)}-\tilde{U}\|_{L^2((0,T);\mathcal{L}(\mathcal{H}^m))}\longrightarrow 0.
\end{aligned}
\end{equation}
Here, the last inequality follows from H\"{o}lder's inequality and the fact that $\sup_{k\ge 1}\|z^{(k)}\|<\infty$. Finally, we have that
\begin{equation}
\begin{aligned}
    \text{(II)}=\text{Re}\bigl[\langle P(0),\tilde{U}(0)-U^{(k)}(0)\rangle_F\bigr]\longrightarrow 0.
\end{aligned}
\end{equation}

Therefore, all terms go to zero, and thus proving weak continuity. 
\end{proof}

\subsection{Fr\'echet derivatives for VQOC}
\label{app:frechetderivatives}
This section provides the derivation of the Fréchet derivatives for VQOC. These derivatives are used in the KKT conditions of Sec. IV of the main text.

\subsubsection*{\textbf{$D_U \mathscr{L}$ derivative}}
Let $\epsilon>0$ and $\delta_U\in \calU$, then

\begin{equation}
\begin{aligned}
    J_1(U+\epsilon \delta_U)-J_1(U)
    =2\epsilon\text{Re}\left(\left\langle \bigl[H_{\text{mol}}|\psi(T)\rangle\langle\psi_0|\bigr],\delta_U(T)\right\rangle_F\right),
    \end{aligned}
    \end{equation}

resulting in

\begin{equation}
    D_U J_1[\delta_U]=\text{Re}\left(\left\langle \bigl(2H_{\text{mol}}|\psi(T)\rangle\langle\psi_0|\bigr),\delta_U(T)\right\rangle_F\right).
\end{equation}

For the constraint term, we have
\begin{equation}
\begin{aligned}
    \langle P,e(U+\epsilon\delta_U,z)\rangle_{\calX,\calX^*}-\langle P,e(U,z)\rangle_{\calX,\calX^*}
    &=\epsilon\int_{0}^T \text{Re}\left(\langle i\partial_t\delta_U(t)-H[z(t)]\delta_U(t), P(t)\rangle_F\right)\,dt,
    \end{aligned}
    \end{equation}
where the second term of the duality product drops out since $\delta_U(0)=0$. Integrating by parts yields
\begin{equation}
\begin{aligned}
     \langle P,e(U+\epsilon\delta_U,z)\rangle_{\calX,\calX^*}-\langle P,e(U,z)\rangle_{\calX,\calX^*}
    &=\epsilon \text{Re}\left(\langle-iP(T),\delta_U(T)\rangle_F-\int_{0}^T\langle -i\partial_tP(t)+ H[z(t)]P(t),\delta_U(t)\rangle_F\,dt\right),
\end{aligned}
\end{equation}

which then results in
\begin{equation}
    D_{U}\langle P,e(U,z)\rangle_{\calX,\calX^*}[\delta_U]=\text{Re}\left(\langle-iP(T),\delta_U(T)\rangle_F-\int_{0}^T\langle -i\partial_tP(t)+ H[z(t)]P(t),\delta_U(t)\rangle_F\,dt\right)
\end{equation}
Since $J_2$ does not depend on $U$, we have in total
\begin{equation}
\begin{aligned}
    D_U \mathscr{L}[\delta_U]&= \text{Re}\left(\bigl\langle\bigl[2H_{\text{mol}}|\psi(T)\rangle\langle\psi_0|-iP(T)\bigr],\delta_U(T)\bigr\rangle_F-\int_{0}^T\langle -i\partial_tP(t)+ H[z(t)]P(t),\delta_U(t)\rangle_F\,dt\right).\\
\end{aligned}
\end{equation}
The KKT conditions then reads
\begin{equation}
 D_U\mathscr{L}(U,z,P)=0 \;\Longleftrightarrow\;
    \left\{\quad\begin{aligned}
     i\partial_tP(t)&=H[z(t)]P(t),\\  P(T)&=-2iH_{\text{mol}}|\psi(T)\rangle\langle\psi_0|,
    \end{aligned}\right.
\end{equation}
which is the adjoint equation. This equation has a terminal condition dependent on $|\psi(T)\rangle$. It can be verified that the solution to this adjoint equation is given in terms of $U(t)$ as
\begin{equation}\label{eq:adjointsolutionapp}
   P(t)=-2iU(t)U^\dagger(T) H_{\text{mol}}|\psi(T)\rangle\langle\psi_0|,
\end{equation}
thus implying that $P\in\calU= W^{1,2}((0,T);\calL(\calH^m))$.

\subsubsection*{\textbf{$D_P \mathscr{L}$ derivative}}
It is not difficult to see that $D_P \mathscr{L}(U,z,P)=0$ is equivalent to $e(U,z)=0$.

\subsubsection*{\textbf{$D_z  \mathscr{L}$ derivative}}
Now for the $z$ derivatives take a perturbation of the form $\epsilon (\tilde z-z)$,  with $\epsilon>0$ and $ \tilde z\in \calZ_{ad}$. Setting $w=\tilde z-z$, we get
\begin{equation}
\begin{aligned}
    J_2(z+\epsilon w)-J_2(z)&=\lambda\sum_{l=1}^L\int_{0}^T\frac{1}{2}(z_l(t)+\epsilon w_l(t))\overline{(z_l(t)+\epsilon w_l(t))}\,dt-\lambda\sum_{l=1}^L\int_{0}^T\frac{1}{2}z_l(t)\overline{z_l(t)}\,dt\\
    &= \epsilon \frac{\lambda}{2} \sum_{l=1}^L\int_{0}^Tz_l(t)\overline{ w_l(t)}+\overline{z_l(t)} w_l(t) \,dt+O(\epsilon^2)=\epsilon \text{Re}\left(\lambda\sum_{l=1}^L\int_{0}^Tz_l(t)\overline{w_l(t)}\,dt\right)+O(\epsilon^2)\\
    &=\epsilon\langle \lambda z, w\rangle_{\calZ}+O(\epsilon^2).
    \end{aligned}
\end{equation}
The derivative then becomes
\begin{equation}
    D_zJ_2[ w]=\langle\lambda z, w\rangle_\calZ.
\end{equation}
For the constraint term many terms will drop out due to linearity in $U$, leaving
\begin{equation}
        \begin{aligned}
     \langle P, e(U,z+\epsilon w)\rangle_{\calX,\calX^*} -\langle P,e(U,z)\rangle_{\calX,\calX^*}
    &=-\epsilon \text{Re}\left(\sum_{l=1}^L\int_{0}^T \Tr\left[\left[\left(  Q_l w_l(t)+Q_l^\dagger\,\overline{ w_l(t)}\right)U(t)\right]^\dagger P(t)\right]\right).
               \end{aligned}
    \end{equation}
    The right-hand side can be further simplified to
\begin{equation}
        \begin{aligned}
    \langle P, e(U,z+\epsilon w)\rangle_{\calX,\calX^*}-\langle P,e(U,z)\rangle_{\calX,\calX^*} 
    &=-\epsilon \sum_{l=1}^L\int_{0}^T\text{Re}\left( \overline{ w_l(t)\Tr\left[Q_lU(t)P(t)^\dagger\right]}+\overline{ w_l(t)}\Tr\left[Q_l^\dagger U(t)P^\dagger(t)\right]\right) \\
    &=\epsilon \left\langle -\Tr\Bigl[Q^\dagger\Bigl(PU^\dagger+U(t)P^\dagger\Bigr)\Bigr], w \right\rangle_{\calZ},
    \end{aligned}
    \end{equation}
where we used the property $\Tr(A^\dagger B)=\overline{\Tr(AB^\dagger)}$ in the first equality and the property $\overline{\Tr(ABC)}=\Tr(C^\dagger B^\dagger A^\dagger)$ in the second.

\medskip

Since $J_1$ is independent of $z$, we have in total
\begin{equation}\label{eq:updateform}
 D_z\mathscr{L}(U,z,P)[\tilde z-z]=\left\langle \lambda z -\Tr\Bigl[Q^\dagger\Bigl(PU^\dagger+UP^\dagger\Bigr)\Bigr], \tilde z - z \right\rangle_{\calZ} \ge 0\qquad\text{for all $\tilde z\in\calZ_{ad}$}.
\end{equation}

\subsection{Parameter Gradient Measurement}\label{app:gradient}
The VQOC algorithm requires evaluation of terms of the form $\langle \psi_1|H|\psi_2\rangle$, where $|\psi_1\rangle\neq|\psi_2\rangle$. A clever trick by Schuld et al.\ in \cite{parametergradient} allows for efficient evaluation of these terms on a quantum computer as long as it is known how to create the $m$-qubit states $|\psi_1\rangle$, $|\psi_2\rangle$, e.g., how to implement $U_1$, $U_2$ s.t. $|\psi_1\rangle =U_1|\psi_0\rangle$ and $|\psi_2\rangle =U_2|\psi_0\rangle.$

\medskip

Note that for any two operators $B$ and $C$, and Hermitian $ Q$, 
\begin{equation}
\begin{aligned}
&\text{Re}\left[\langle\psi|B^{\dagger}  Q C| \psi\rangle\right]=\frac{1}{2}\langle\psi|B^{\dagger}  Q C| \psi\rangle+\mathrm{h.c.}= \frac{1}{4}\left[\langle\psi|(B+C)^{\dagger}  Q(B+C)| \psi\rangle-\langle\psi|(B-C)^{\dagger}  Q(B-C)| \psi\rangle\right],\\
&\text{Im}\left[\langle\psi|B^{\dagger}  Q C| \psi\rangle\right]=\frac{1}{2}\langle\psi|B^{\dagger}  Q C| \psi\rangle-\mathrm{h.c.}= \frac{i}{4}\left[\langle\psi|(B-iC)^{\dagger}  Q(B-iC)| \psi\rangle-\langle\psi|(B+iC)^{\dagger}  Q(B+iC)| \psi\rangle\right],
\end{aligned}
\end{equation}

where h.c.\ refers to the Hermitian conjugate. Figure~\ref{fig:measurementqoc} shows the quantum circuit diagram of the scheme for determining $\text{Re}[\langle\psi_0|U_1^\dagger HU_2|\psi_0\rangle]$. In order to determine $\text{Re}[\langle\psi_0|U_1^\dagger HU_2|\psi_0\rangle]$ the product state of the initial state $|\psi_0\rangle$ with an ancilla qubit $|0\rangle$ is considered. A Hadamard gate is applied to the ancilla qubit to obtain $(|0\rangle+|1\rangle)/\sqrt{2}\otimes|\psi_0\rangle$.  $U_1$ is applied to $|\psi_0\rangle$ conditioned on the fact that the ancilla is $|0\rangle$, and $U_2$ is applied to $|\psi_0\rangle$ conditioned on the fact that the ancilla is $|1\rangle$. The product state then becomes entangled as
\begin{equation}
\frac{1}{\sqrt{2}}\Bigl(|0\rangle \otimes U_1|\psi_0\rangle+|1\rangle \otimes U_2|\psi_0\rangle\Bigr).
\end{equation}
Another Hadamard gate is applied to the ancilla qubit to get the final state
\begin{equation}
\frac{1}{2}\Bigl(|0\rangle\otimes(U_1+U_2)|\psi_0\rangle+|1\rangle\otimes(U_1-U_2)|\psi_0\rangle\Bigr).
\end{equation}

\begin{figure}[H]
\begin{center}
\begin{tikzcd}
\lstick{$\ket{0}$} & \gate{H}  & \octrl{1} &\ctrl{1} & \gate{H} &\meter{} & \cw & \lstick{c} \\
\lstick{$\ket{\psi}$} & \qw & \gate{U_1} & \gate{U_2} & \qw &\qw & \qw & &\lstick{$\ket{\psi'}$}
\end{tikzcd}
\end{center}
    \caption{Quantum circuit diagram of the parameter gradient scheme for determining $\text{Re}[\langle\psi_0|U_1HU_2|\psi_0\rangle]$}
    \label{fig:measurementqoc}
\end{figure}

A measurement on the ancilla results in either 
\begin{equation}
    \begin{aligned}
    &\left|\psi_{0}^{\prime}\right\rangle=\frac{1}{2 \sqrt{p_{0}}}(U_1+U_2)|\psi\rangle,\quad\quad p_{0}=\frac{1}{4}\left\langle\psi\left|(U_1+U_2)^{\dagger}(U_1+U_2)\right| \psi\right\rangle,\\
    &\left|\psi_{1}^{\prime}\right\rangle=\frac{1}{2 \sqrt{p_{1}}}(U_1-U_2)|\psi\rangle,\quad\quad p_{1}=\frac{1}{4}\left\langle\psi\left|(U_1-U_2)^{\dagger}(U_1-U_2)\right| \psi\right\rangle,
    \end{aligned}
\end{equation}
with $p_0, p_1$ the probabilities of measuring these states. Repeating this procedure allows for estimation of $p_0$ and $p_1$. The expectation of $H$ is then measured on these states to get
\begin{equation}
\begin{aligned}
    &\tilde{E}_{0}=\left\langle\psi_{0}^{\prime}|H| \psi_{0}^{\prime}\right\rangle=\frac{1}{4 p_{0}}\left\langle\psi\left|(U_1+U_2)^{\dagger} H(U_1+U_2)\right| \psi\right\rangle,\\
    &\tilde{E}_{1}=\left\langle\psi_{1}^{\prime}|H| \psi_{1}^{\prime}\right\rangle=\frac{1}{4 p_{1}}\left\langle\psi\left|(U_1-U_2)^{\dagger} H(U_1-U_2)\right| \psi\right\rangle,
    \end{aligned}
\end{equation}
and therefore
\[
    \text{Re}\left[\langle\psi_0|U_1^\dagger HU_2|\psi_0\rangle\right]=\frac{1}{2}\left\langle\psi\left|U_1^{\dagger} H U_2\right| \psi\right\rangle+\mathrm{h.c.}=\left(p_{0} \tilde{E}_{0}-p_{1} \tilde{E}_{1}\right).
\]
The imaginary part can be found analogously by applying $iU_2$, instead of $U_2$, conditional on the ancilla qubit being 1 instead of applying $U_2$. By linearity of the inner product, any term of the form
\begin{equation}
\left\langle\psi_0\left|\left[\sum_{k=1}^{K_1} U_{1,k}^\dagger\right] H\left[\sum_{k=1}^{K_2}U_{2,k}\right]\right|\psi_0\right\rangle,
\label{eq:schuldresult}
\end{equation}
can be efficiently evaluated, as long as it is possible to apply $U_{i,k}$ to an $m$-qubit state.

\medskip

One of the things not mentioned in the original paper by Schuld et al.\ \cite{parametergradient} is how to create controlled gates for random unitaries. One way to create a controlled-$U$ gate, for an arbitrary one qubit unitary $U$, is to write $U=e^{-i\alpha}AXBXC$, which is always possible \cite{onequbitdecomop}, and apply the circuit, with CNOT gates, as in Fig.~\ref{fig:cnotpairs}.

\begin{figure}[H]
\begin{center}
\begin{tikzcd}
\\
\lstick{$\ket{q_1}$} & \ctrl{1} & \qw \\
\lstick{$\ket{\psi}$} & \gate{U} & \qw 
\end{tikzcd}
=
\begin{tikzcd}
\lstick{$\ket{q_1}$} & \qw  & \ctrl{1} &\qw &\ctrl{1} & \gate{|0\rangle\langle0|+e^{-i\alpha}|1\rangle\langle1|} &\qw \\
\lstick{$\ket{\psi}$} & \gate{A} & \targ{2} & \gate{B} &\targ{2} & \gate{C} & \qw
\end{tikzcd}
\end{center}
\caption{Creating a controlled $U$ gate using 1 qubit control.}
\label{fig:cnotpairs}
\end{figure}

$m$-qubit gates $U$ can be decomposed into Pauli operators and applied as individual controlled gates to the qubits, for example,
\begin{equation}
    |00\rangle\langle00|+|01\rangle\langle10|+|10\rangle\langle01|+|11\rangle\langle11|=\left(\begin{array}{cccc} \
1 & 0 & 0 & 0 \\ 0 & 0 & 1 & 0 \\ 0 & 1 & 0 & 0 \\ 0 & 0 & 0 & 1 \end{array}\right)=\frac{1}{2}(II+ZZ+XX+YY).
\end{equation}

In the VQOC algorithm, the gradients as in Eq.~\eqref{eq:updateform} need to be  determined on a quantum computer. It is easiest to write out the update form in state form instead of matrix from as
\begin{equation}
\begin{aligned}
    &\Tr[Q_l^\dagger(P(t)U^\dagger(t)+U(t)P^\dagger(t))]\\
    &\hspace{4em}=\Tr[Q_l^\dagger(-2iU(t)U^\dagger(T) H_{\text{mol}}|\psi(T)\rangle\langle\psi_0|U^\dagger(t)+2i|\psi(t)\rangle\langle\psi_0| U^\dagger(T) H_{\text{mol}} U(T)U^\dagger(t))]\\
    &\hspace{4em}=-2i(\langle \psi_0|U^\dagger(t) Q_l^\dagger U(t)U^\dagger(T) H_{\text{mol}} U(T) |\psi_0\rangle-\langle\psi_0|U^\dagger(T) H_{\text{mol}} U(T)U^\dagger(t) Q_l^\dagger U(t)|\psi_0\rangle).
    \label{eq:470}
\end{aligned}
\end{equation}
The propagators $U(T)U^\dagger(t)$ can be easily applied by splitting $U(T)$ as $U(T)=\Gamma(t,T)U(t)$, where $\Gamma(t,s)=U(t)U^\dagger(s)$ can be interpreted as the evolution operator resulting from applying the pulse in the interval $[t,s).$ As discussed in Sec. IV of the main text the control Hamiltonians $Q_l$ normally decompose in a small number of single qubit Pauli operators which can be applied quickly to the system, e.g. $Q_l=\sum_{k=1}^K V_{l,k}$. Because of this, it is possible to apply the unitary $\Gamma(t,T) V_{l,k} U(t)$ to a state on a Rydberg quantum computer by applying the pulse up to $t$, applying $V_{l,k}$ in a relatively short time, and then applying the rest of the pulse up to $T$. The individual terms in Eq.~\eqref{eq:470} can be written in the form
\begin{equation}
\begin{aligned}
    \left\langle \psi_0\left|\sum_{k=1}^K U^\dagger(t) V_{k,l}^\dagger U(t)U^\dagger(T) H_{\text{mol}} U(T) \right|\psi_0\right\rangle
    =\sum_{k=1}^K\langle\psi(t)|\Gamma^\dagger(t,T)  H_{\text{mol}} \Gamma(t,T) V_{k,l}^\dagger |\psi(t)\rangle.
\end{aligned}
\end{equation}
By identifying $|\psi_0\rangle=|\psi_t\rangle$, $U_{1,k}=V_{k,l}$, $U_{2,k}=I$ and $H=\Gamma^\dagger(t,T)  H_{\text{mol}} \Gamma(t,T)$ as in Eq.~\eqref{eq:schuldresult}, the parameter gradient algorithm can be applied  to determine these terms efficiently on a quantum computer. $H$ can be evaluated for a qubit state by applying $\Gamma(t,T)$ to the output states of the parameter gradient algorithm and then evaluating $H_{\text{mol}}.$

\medskip

Since both terms in Eq.~\eqref{eq:470} have to be evaluated at each time step and for each value of $L$ the number of quantum evaluations that needs to be done is given by
\begin{equation}
    \#QE=2\cdot K\cdot L\cdot N\cdot \#\text{shots}.
    \label{eq:costQOC}
\end{equation}

For the parameter gradient algorithm to be implementable, it needs to be checked if $U_{1,k}$ and $U_{2,k}$ are easily implementable in order to create controlled gates. In the case of $U_{2,k}=I$ this is trivial and  $U_{1,k}=V_{k,l}$ is a simple one or two qubit unitary for which the decomposition can be found.

\medskip

\bibliographystyle{apsrev4-1}
\bibliography{Bibliography2}